\documentclass{CUP-JNL-DCE}
\AtBeginDocument{%
  \setlength{\textfloatsep}{6pt}
  \setlength{\intextsep}{6pt plus 2pt minus 2pt}
  \setlength{\abovecaptionskip}{4pt}
  \setlength{\dbltextfloatsep}{4pt plus 2pt minus 2pt}
}
\usepackage{latexsym}
\usepackage{graphicx}
\usepackage{fix-cm}
\usepackage{multicol,multirow}
\usepackage{amsmath,amssymb,amsfonts}
\usepackage{mathrsfs}
\let\mathscr\mathcal
\usepackage{amsthm}
\usepackage{rotating}
\usepackage{appendix}
\usepackage{ifpdf}
\usepackage[T1]{fontenc}
\usepackage{times}
\usepackage{textcomp}
\usepackage{xcolor}
\usepackage{hyperref}
\usepackage{lipsum}
\usepackage{booktabs}
\usepackage{enumitem}
\usepackage{indentfirst}          
\usepackage{tabularx}
\theoremstyle{definition}

\usepackage{longtable}
\usepackage{booktabs}
\usepackage{array}
\usepackage{algorithm}
\usepackage{algpseudocode}
\usepackage[numbers,sort&compress]{natbib}
\usepackage{cleveref}

\usepackage{draftwatermark}
\usepackage{fancyhdr}

\SetWatermarkText{PREPRINT}
\SetWatermarkScale{1}      
\SetWatermarkColor[gray]{0.95} 

\articletype{DATA ARTICLE}
\jname{Data-Centric Engineering}
\jyear{2026}
\begin{document}
\setlength{\headheight}{12pt}
\setlength{\footskip}{12pt}
\begin{Frontmatter}
\title[Article Title]{Numerical benchmark for damage identification in Structural Health Monitoring}
\author*[1]{Francesca Marafini}\email{francesca.marafini@unifi.it}\orcid{0000-0002-8383-8508}
\author[1]{Giacomo Zini}
\author[2]{Alberto Barontini}
\author[3]{Nuno Mendes}
\author[4]{Alice Cicirello}
\author[1]{Michele Betti}
\author[1]{Gianni Bartoli}
\authormark{Francesca Marafini \textit{et al}.}
\address*[1]{\orgdiv{Department of Civil and Environmental Engineering (DICEA)}, \orgname{University of Florence}, \orgaddress{\city{Florence},  \country{Italy}}}
\address[2]{\orgdiv{Department of Engineering and Geology (InGeo)}, \orgname{University of Chieti-Pescara}, \orgaddress{\city{Pescara},  \country{Italy}}}
\address[3]{\orgdiv{Department of Civil Engineering}, \orgname{University of Minho}, \orgaddress{\city{Guimarães},  \country{Portugal}}}
\address[4]{\orgdiv{Department of Engineering}, \orgname{University of Cambridge}, \orgaddress{\city{Cambridge}, \country{UK}}}
\received{04 March 2026}
\revised{28 September 2026}
\accepted{-}
\abstract{The availability of a dataset for validation and verification purposes of novel data-driven strategies and/or hybrid physics-data approaches is currently one of the most pressing challenges in the engineering field, constraining the controlled evaluation and comparison of SHM methods.
Data ownership, security, access and metadata handiness are currently hindering advances across many fields, particularly in Structural Health Monitoring (SHM) applications. This paper presents a simulated SHM dataset, comprised of dynamic and static measurements (i.e., acceleration and displacement), and includes the conceptual framework designed to generate it. The simulated measurements were generated to incorporate the effects of Environmental and Operational Variations (EOVs), different types of damage, measurement noise and sensor faults and malfunctions, in order to account for scenarios that may occur during real acquisitions. A fixed-fixed steel beam structure was chosen as reference for the numerical benchmark. The simulated monitoring was operated under the assumptions of a Single Degree of Freedom (SDOF) for generating acceleration records and of the Euler-Bernoulli beam for the simulated displacement measurements. The generation process involved the use of parallel computation, and together with the generated data is available open-source, ensuring reproducibility, repeatability and accessibility for further research. Together, the openly available dataset and generation code provide a valuable resource for researchers aiming to develop, refine or compare SHM methods under controlled combinations of environmental, operational, structural, and measurement-related changes.}

\begin{policy}[Impact Statement]
The proposed benchmark is designed to support the comparable evaluation of SHM methods in a fully defined, controllable environment. It enables the possibility of systematic testing through the assessment of sensitivity to imposed structural changes separately from sensitivity to the other simulated effects, also called confounding influences, which in real data could mimic or mask the presence of damage. In addition, it allows for the quantification of the robustness to sensor faults and malfunctions, and the repetition of ablation studies where a single effect can be included or excluded in the data simulation. Paired dynamic and static recordings also support heterogeneous data fusion and consistency checks between response types. Since the benchmark is parametrized and reproducible, it can serve as a common reference point for reporting performance and uncertainty, facilitating fairer comparison between data-driven, physics-based and hybrid approaches, and accelerating the adaptation of specific damage identification methods moving from numerical to experimental and real-world testing.
\end{policy}

\end{Frontmatter}

\section{Introduction}
\label{sec1}
Structural Health Monitoring (SHM) methods have been increasingly applied to smart management of structures and infrastructures, with the aim of preventive cost-effective preservation of critical assets. Although many tools for monitoring, data analysis, and prediction have become available in the last decades, the validation and testing of their performance against large realistic datasets is still a challenging task due to the reduced number and limitations of the existing open access benchmarks. Real data presents intrinsic variations caused by Environmental and Operational Variabilities (EOVs), such as temperature fluctuations and changes in use. Moreover, monitoring data may be affected by additional sources of uncertainties and errors such as sensor faults or malfunctions. These factors, combined and sometimes interdependent, can hinder the interpretability of the results and mask the effects of damage in the short-term (or fast-varying damage, causing sudden drops of performance in the short-term) and in the long-term (slow-varying damage, for instance due to the natural ageing of the structure over time). To evaluate SHM techniques under controlled conditions, numerous experimental benchmarks have been developed, such as the well-known Z24 bridge \cite{maeckDescriptionZ24Benchmark2003} and Hawk T1A aircraft \cite{haywood-alexanderFullscaleModalTesting2024}. The time and economic constraints of damaging real prototypes needed to generate large amounts of labelled data have recently led to the adoption of numerical benchmarks. These often start from simple mechanical system formulations \cite{wordenStructuralFaultDetection1997} or expand existing experimental setups \cite{svendsenDatabasedStructuralHealth2022}, focusing mainly on simulations of damage scenarios. Some of these studies have integrated environmental variabilities \cite{tatsisNumericalBenchmarkSystem2019} and long-term damage effects \cite{tatsisVibrationbasedMonitoringSmallscale2021a}, but only in specific cases and never in combination with sensor faults. An account of recent numerical benchmark studies available today in the literature is provided in Table~\ref{tab:Table1}. For studies in Table~\ref{tab:Table1} that are  starred*, the data are also experimental, while the symbol $^{\dagger}$ is also present in the apex, the data or the model to generate them is available for public use. This paper presents a synthetically generated dataset designed to provide controlled, labelled SHM data combining selected representations of environmental and operational variability, sudden and progressive structural changes, and sensor faults and malfunctions. The resulting dataset and the code written to generate it were made available in open source, for testing and benchmarking damage identification methods, addressing key open challenges in synthetic SHM data generation.

\begin{table}[t]
\caption{Overview of published numerical SHM benchmarks}
\label{tab:Table1}
\centering
\setlength{\tabcolsep}{4pt}
\renewcommand{\arraystretch}{1.15}
\begin{tabular}{
p{0.05\linewidth}
p{0.05\linewidth}
>{\centering\arraybackslash}p{0.19\linewidth}
>{\centering\arraybackslash}p{0.19\linewidth}
>{\centering\arraybackslash}p{0.19\linewidth}
>{\centering\arraybackslash}p{0.19\linewidth}
}

\toprule
\textbf{Ref.} & \textbf{Year} & \textbf{Structure} & \textbf{Objective} & \textbf{OVs} & \textbf{FAST} \\
\midrule

~\cite{wordenStructuralFaultDetection1997} & 1996 &
Multi Degree of Freedom (MDOF) mechanical system &
Novelty detection using the transmissibility function as a feature &
Harmonic excitation &
1, 10 and 50\% stiffness reduction \\

~\cite{spencerNextGenerationBenchmark1999}$^{*}$& 1999 &
SAC Phase II Steel Moment Frame &
Test control strategies after strong motion earthquakes &
Ground acceleration &
Stiffness reduction corresponding to an 18\% reduction in frequency \\

~\cite{johnsonPhaseIASCASCEStructural2004}$^{\dagger}$ & 2002 &
IASCE--ASCE SHM 2x2bay steel frame &
Evaluation of SHM methods &
White Gaussian Noise (WGN), Shakers &
6 scenarios with braces removal or bolts loosening \\

~\cite{burkettBenchmarkStudiesStructural2005}$^{*\dagger}$ & 2003 &
IABMAS 1--2 span bridge &
Test damage detection algorithms &
WGN, Hammer impact, Shakers &
Reduced stiffness, Boundary condition change \\

~\cite{tisoNewChallengingBenchmark2017}$^{\dagger}$ & 2017 &
Two offset cantilevered beams &
Complex nonlinear dynamics for system identification &
Load amplitude variability &
No damage scenarios \\

~\cite{tatsisNumericalBenchmarkSystem2019}$^{\dagger}$ & 2019 &
2-span steel beam &
Validation of decision-making tools &
WGN, Deterministic moving load &
6 scenarios with stiffness reduction \\

~\cite{tatsisVibrationbasedMonitoringSmallscale2021a}$^{\dagger}$ & 2021 &
Wind turbine blade &
Condition assessment of a wind turbine blade &
Rotor speed changes &
Cracks openings simulated as local stiffness reduction \\

~\cite{vlachasTwostoryFrameBoucWen}$^{\dagger}$ & 2021 &
Two-story steel frame structure with Bouc--Wen hysteretic links &
Validations of SHM, model reduction, and structural identification methods &
Ground motion excitation scenarios &
Strength and stiffness deterioration Bouc--Wen model parameter variations  \\

~\cite{svendsenDatabasedStructuralHealth2022}$^{*}$ & 2022 &
Steel truss bridge (Hell Bridge Test Arena) &
Hybrid SHM for DI on steel bridges &
WGN &
Stiffness reduction in connections \\

\bottomrule
\end{tabular}
\end{table}

\section{Methodology}
\label{sec2}
\subsection{Framework}
\begin{figure}[!h]
    \FIG{\includegraphics[width = 1\linewidth]{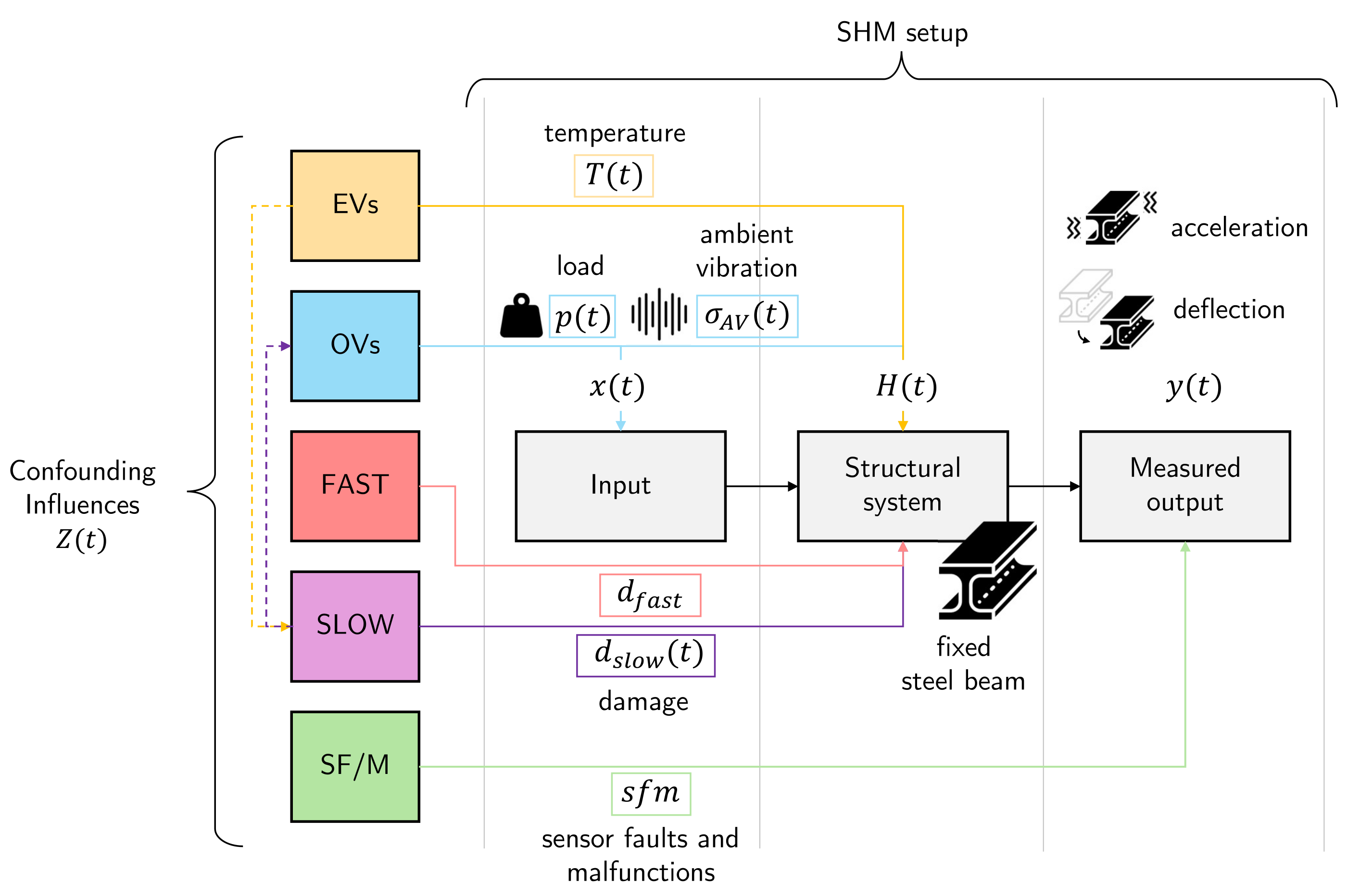}}
    {\caption{Chosen simplified SHM setup representation for the benchmark and schematic representation of the modelled confounding influences Z(t) acting on the SHM system}
    \label{fig:Figure1}}
\end{figure}

The numerical generation of SHM data was designed using a straightforward and scalable approach, based on a simple but informative case study, to combine controlled variation of selected SHM influences with computationally efficient multi-year simulation. A generic SHM setup was considered, in which three main elements can be identified: a structural system $H(t)$ with time-varying properties, an input excitation $x(t)$, and a measured response $y(t)$ resulting from the transfer of $x(t)$ through $H(t)$ (see the sketch in Fig.~\ref{fig:Figure1}). In real data, numerous influences (denoted as $Z(t))$ are observed affecting each of the three components, such as changes in the system properties (e.g., temperature-dependent material variations), or in the nature of input (e.g., excitation variability), or in the acquisition conditions (e.g., sensor drift or electronic degradation) and be captured by the measured response. Such effects are commonly referred to as confounding influences as they can mask or mimic damage in the measured response. In this work, confounding influences were modelled as time-varying covariate vectors affecting the system, the input, or the measured response. In this context, variability is intended to denote parameter fluctuations (like operational (OV) and environmental variability (EV)), while variation refers to the resulting changes in structural response (such as frequency shifts due to stiffness variations).\par
Although a beam was used here as a specific case study, the benchmark generation procedure is general and could be extended to other structural configurations and additional sources of influence. The framework described here provides a reproducible procedure for combining prescribed environmental, operational, structural, and measurement-related changes within a common monitoring timeline. Individual modelling components could be refined or replaced, provided that their dependencies and their effects on the generated responses are consistently reformulated.\par
The system was assumed to be linear but time-variant over the monitoring period. A single-input single-output configuration was considered, with one monitored Degree Of Freedom (DOF). As a reference case, a fixed-fixed steel beam subjected to varying static loads and ambient vibration excitation was selected. The beam was monitored through deflection measurements to capture its static response and acceleration signals to capture its dynamic response. \par  Among all possible $Z(t)$ that could interact to shape SHM data, a selection of them was explicitly modelled, in order to reproduce real-world phenomena either directly (e.g., modelling the dependence of the Young’s modulus of steel on temperature) or indirectly (e.g., simulating sensor faults a posteriori on the generated data).\par

The following effects were included: 
\begin{itemize}[leftmargin = 1.5em]
  \item the variability of the operational static load $p$, including short- and long-term components;
  \item the variability of the ambient vibration within a fixed frequency band, with varying amplitude $\sigma_{AV}$;
  \item the effect of the variability of the environment on material properties, modelled through an empirical relationship between steel stiffness and temperature $T$;
  \item the effect of sudden damage (FAST), appearing as a shift in the data, modelled as a decrease in stiffness $d_{\mathrm{fast}}$;
  \item the effect of progressive damage (SLOW), simulating progressive structural changes through a gradual reduction of the beam area, stiffness, and mass $d_{\mathrm{slow}}$;
  \item the effect of system's sensor faults and malfunctions (SF/M) applied a posteriori to the simulated data ($sfm$).
\end{itemize}
The simulated effects qualitatively illustrated in Fig.~\ref{fig:Figure1} were simulated to act either on the input, the parameters of the structural system, or the measured output within the design setup. The dotted lines highlight cases in which confounding influences were simulated to act indirectly, for example, for environmental parameters that govern slow damage evolution or for the same evolution affecting operational conditions. The simulated monitoring covered a time span of three years, enabling the representation of both slow processes (seasonal cycles, long-term trends) and intermittent or sudden events (operational changes, fast damage, sensor faults). For static monitoring, single value hourly acquisitions of midspan deflection were generated to simulate a long-term displacement record, while for dynamic monitoring, the synthetic generated data consisted of 3-minute acceleration time histories simulating AV acquisitions. Within this framework, two models were adopted to compute the system response: a Euler-Bernoulli beam model for the static response and a linear single-degree-of-freedom (SDOF) model for the dynamic response. \par
 
\begin{figure}[!h]
  \includegraphics[width = 1\linewidth]{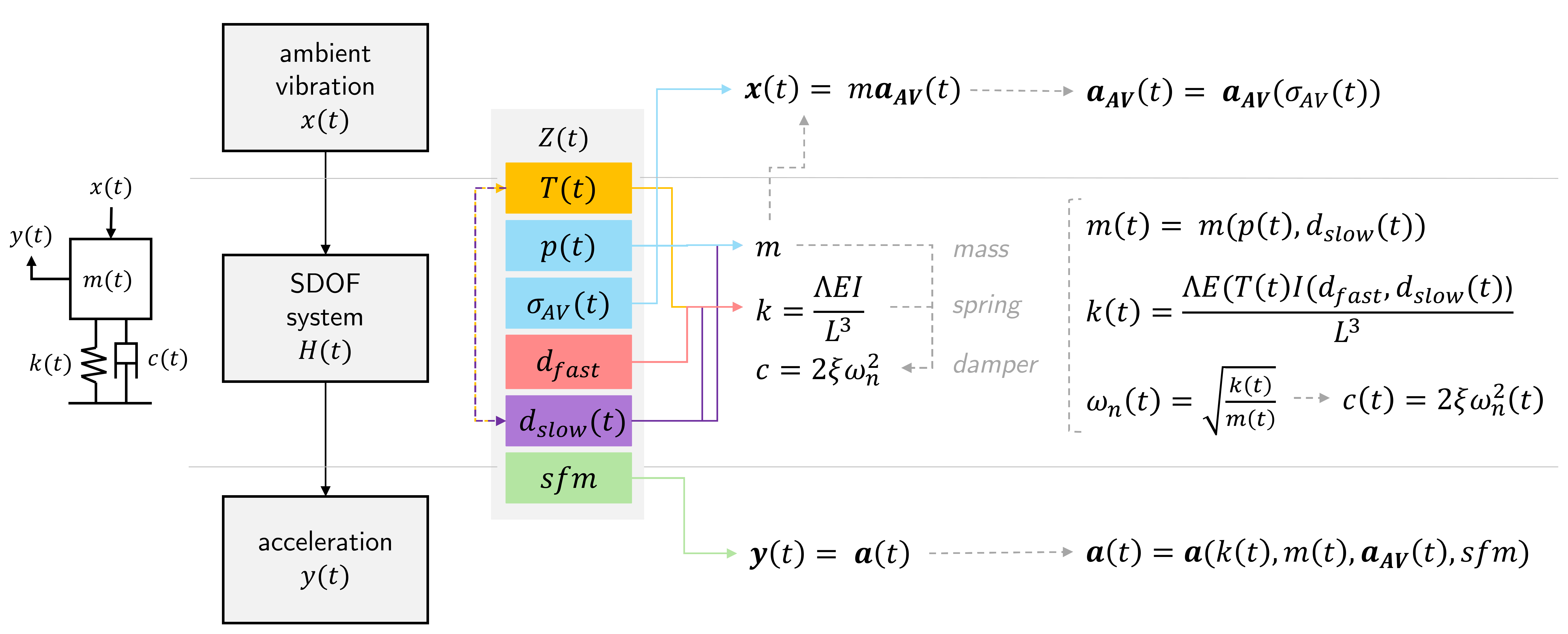}
  \caption{SDOF formulation including the confounding influences}
  \label{fig:Figure2}
\end{figure}

In the formulations reported in Figure~\ref{fig:Figure2}  for the simulation of the dynamic response, the SDOF stiffness considered was the beam bending stiffness $k$ and depended on the following: the boundary conditions $\Lambda$; the elastic modulus $E$, which in turn depended on the temperature varying with time $T(t)$ and the section inertia $I$ affected by FAST and SLOW varying damages, due to applied sudden stiffness drops and progressive reduction of the cross-section area (damage simulations are described in Sections~3.6.3.1 and~3.6.3.2). The mass $m$ instead was not considered affected by EVs, but only by the variability of the live load considering a slow application rate, and slightly by SLOW damage, due to the loss of volume and consequently the reduction of the cross-section. The variability in time of the ambient excitation  $\sigma_{AV}$ was applied to the input, while the SF/M were instead only applied to the measured output. The output acceleration depends on all of the above parameters, and the generalized equation of motion can be expressed as follows in Eq.~\ref{eq:eq0}.

\begin{equation}
	\begin{aligned}
		m\mathbf{a}(t) + c\mathbf{v}(t) + k\mathbf{u}(t)
		&= m\mathbf{a}_{\mathrm{AV}}(t).
		\label{eq:eq0}
	\end{aligned}
\end{equation}

\begin{figure}[!h]
  \includegraphics[width = 1\linewidth]{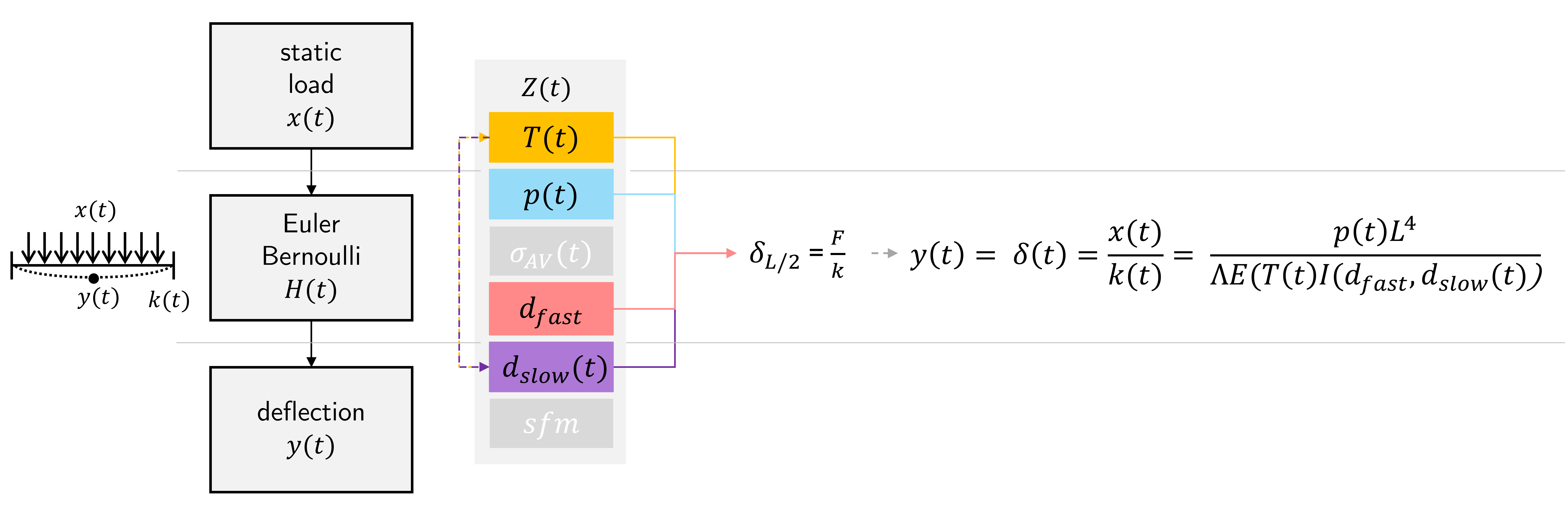}
  \caption{Midspan deflection calculation with Euler-Bernoulli including the confounding influences}
  \label{fig:Figure3}
\end{figure}

By contrast, the deflection was calculated at midspan ($L/2$) according to Euler-Bernoulli, proportionally to the applied static load (OVs) and inversely to the beam bending stiffness, therefore depending on EVs (affecting the Young Modulus) and FAST and SLOW varying damage affecting $I$ (Figure~\ref{fig:Figure3}).\par

Moving from the time-dependent notation used in Figures~\ref{fig:Figure2} and ~\ref{fig:Figure3}, the cascading effect of the confounding influences was expressed by discretising the continuous-time formulations. Each acquisition was generated considering finite hourly instants at which the simulated effects, expressed as covariate parameters, were assumed constant (i.e.\ temperature was not considered to vary within a single 3-minute acceleration signal). Therefore, the discrete sequences of hourly acquisitions, generated over the 3-year monitoring period, were indexed by $i = 1,\dots,n_a$ ($n_a$ = number of acquisitions). Each dynamic acquisition was denoted as one acceleration vector $\mathbf{a}_i\in\mathbb{R}^{n_s}$, where $n_s$ is the number of samples within the acquisition. Each simulation of static measurement was denoted as the scalar $\delta\in\mathbb{R}^1$.

For each hourly acquisition $i$, the benchmark data generation was setup to simulate the cumulative effect of the variability of temperature $T_i$, of the operational load $p_i$, and two types of damage effects: $d_{\mathrm{FAST},i}$ for sudden events (e.g., stiffness decay indicative of rapid damage occurrence), and $d_{\mathrm{SLOW},i}$ for progressive damage (e.g., corrosion), and of sensor malfunction or failures denoted as $\mathrm{sfm}_i$, capturing the anomalies in data not directly related to the structural health itself, applied only to the generated accelerations $a_i$ a posteriori. The term $\varepsilon_i$ was also included to possibly account for potential random errors or noise within acquisition $i$, modelled as White Gaussian Noise (WGN). All these effects were synthetically expressed in vectors of covariates at each acquisition level as shown in Eq.~\ref{eq:eq1} and Eq.~\ref{eq:eq2} which show the formalisation of the qualitative models describing how each acquisition depends on its covariates. The cascading effect of the confounding influences was also expressed by discretizing the continuous-time formulations. Each acquisition was generated considering finite time hourly instants at which the simulated effects, expressed as covariate parameters, were assumed constant (i.e. temperature is not considered to vary within a single 3-minute acceleration signal). Therefore, the discrete sequences of hourly acquisitions, generated over the 3-year monitoring period, were indexed by $i\ =\ 1,\ldots,\ n_a$ ($n_a$ = number of acquisitions). Each dynamic acquisition was denoted as one acceleration vector $\mathbf{a}_i \in \mathbb{R}^{n_s}$, where $n_s$ is the number of samples within the acquisition. Each simulation of static measurement was denoted as the scalar $\delta\in\mathbb{R}^1$. 

\begin{equation}
\begin{aligned}
\mathbf{z}_i^{(\mathrm{dyn})} &= \bigl[ T_i,\; p_i,\; \sigma_{\mathrm{AV},i},\; d_{\mathrm{fast},i},\; d_{\mathrm{slow},i},\; sfm_i \bigr], \\
\mathbf{z}_i^{(\mathrm{stat})} &= \bigl[ T_i,\; p_i,\; d_{\mathrm{fast},i},\; d_{\mathrm{slow},i} \bigr].
\label{eq:eq1}
\end{aligned}
\end{equation}

\begin{equation}
\begin{aligned}
a_i &= f\!\left(\mathbf{z}_i^{(\mathrm{dyn})}\right) + \varepsilon_i, \\
\delta_i &= f\!\left(\mathbf{z}_i^{(\mathrm{stat})}\right) + \varepsilon_i.
\label{eq:eq2}
\end{aligned}
\end{equation}

\subsection{Applicability and limitations}

The synthetic data generation simulates acceleration time-histories, displacement data, and assumes temperature measurements. The resulting benchmark is intended as a controlled testing ground for SHM methods rather than as a realistic numerical representation of a specific real structure. Its main objective is to provide multi-year, fully labelled data in which EOVs, structural changes with different temporal evolution, and SF/M can be introduced independently or in combination, so that their individual effects can be isolated when needed and their interaction can be investigated under known conditions. This type of controlled setup is particularly useful in long-term SHM, where the distinction between structural changes and variations induced by environmental, operational, or acquisition-related effects remains one of the main challenges. Therefore, the present work also responds to the limited availability of long-term labelled datasets in which EOVs, progressive or sudden structural changes, and measurement anomalies coexist with known ground truth. Reproducing equivalent combinations experimentally over multi-year monitoring periods would require extensive monitoring campaigns and, particularly for known damage progression and sensor-fault conditions, controlled interventions that are difficult to implement systematically. The numerical benchmark provides a starting point in which these challenges can be investigated under defined assumptions before introducing additional physical and experimental complexity.

A wide range of methods relying on single-channel or global response information can be developed, compared, and evaluated on this data, including:
	\begin{itemize}
		\item extraction, selection, and comparison of Damage-Sensitive Features (DSFs);
		\item assessment of feature sensitivity to environmental, operational, damage-related, and measurement-related variability;
		\item compensation, normalisation, or disentanglement of Environmental and Operational Variabilities (EOVs);
		\item investigation of confounding effects, including cases in which EOVs, structural changes, and sensor faults produce partially overlapping changes in the measured response;
		\item novelty and anomaly detection under undamaged reference conditions;
		\item detection and identification of sudden changes in structural response;
		\item detection of progressive or slowly evolving structural changes and long-term trend identification;
		\item comparison of detection thresholds and minimum detectable structural changes;
		\item sensor-fault detection, classification, and discrimination between structural anomalies and measurement anomalies;
		\item development of statistical, Machine Learning (ML), Deep Learning (DL), and Physics-Enhanced approaches;
		\item heterogeneous data-fusion approaches combining static and dynamic response quantities.
	\end{itemize}

The adopted data generation approach also defines the range of applications for which the benchmark is not intended as the equivalent SDOF representation provides only global dynamic information and does not reproduce the spatial response of the beam or multiple structural modes. Consequently, it is not suitable for approaches requiring an array of sensors, relative information between measurement locations, mode shapes or mode-shape curvature, modal flexibility, or other spatial quantities typically used for damage localisation. Damage is represented through changes in equivalent structural properties, rather than through a detailed description of the local physical mechanisms producing those changes. The benchmark can therefore be used to investigate whether, when, and to what extent a change in structural condition can be identified from the measured global response, while the localisation of damage and the identification of its specific physical mechanism remain outside its scope.
The structural response is represented through linear Euler-Bernoulli and SDOF formulations. Nonlinearities associated with material behaviour, geometry, constraints, contact or slip, crack opening and closing, or other amplitude-dependent mechanisms are not included. The dataset does not support evaluation of methods whose target behaviour requires structural nonlinearities, but at the same time it allows for the investigation of the effects of EOVs, structural changes, and measurement-related disturbances without introducing nonlinear structural behaviour as an additional complexity. The benchmark should therefore be interpreted within the range of low-amplitude operational conditions for which an equivalent linear representation is adopted.
Environmental and operational variability are represented through a selected set of time-varying quantities. Environmental effects are not modelled spatially and coupled thermo-mechanical phenomena are not represented, while the operational model does not reproduce the full spatial and temporal complexity of real loading and excitation conditions. Additional source-specific excitation mechanisms were not explicitly modelled. The purpose of these modelling choices is to introduce controlled variations capable of modifying the structural response while maintaining a known relationship between the imposed conditions and the generated data.
The same principle was adopted for sensor faults and malfunctions, which are represented according to their effects on the measured signals rather than through complete physical models of the sensing and acquisition system. This allows measurement-related anomalies to be introduced independently of changes in structural condition, while mechanisms that depend on a specific sensor technology, mounting configuration, acquisition chain, or sensor-structure interaction are not explicitly represented.
Uncertainty was treated probabilistically for the live-load process and random sampling was also used in excitation generation and fault assignment, but no uncertainties were propagated in geometry, material properties, and constraints. Randomness was nevertheless introduced in other components of the data-generation process, as described in the following sections. Several reference properties were intentionally kept deterministic in order to preserve a clearly defined ground truth and to make the contribution of the different simulated effects identifiable and reproducible.

The main modelling choices, together with their limitations and advantages for the applicability of the benchmark, are summarised in Table~\ref{tab:Table2bis}. These choices represent deliberate trade-offs between physical complexity, controllability, and computational cost. The benchmark is not intended to reproduce all the phenomena that may occur in a real SHM system, nor to provide a one-to-one numerical counterpart of a particular monitored structure, but to offer a common and reproducible baseline in which several challenges of long-term SHM can first be investigated under controlled assumptions, a baseline that can be extended as the open-source code offers opportunities for future expansions.

\begin{table*}[t]
	\centering
	\caption{Summary of the main modelling choices, limitations, and advantages of the benchmark.}
	\label{tab:Table2bis}
	
	\renewcommand{\arraystretch}{1.15}
	
	\begin{tabularx}{\textwidth}{
			>{\raggedright\arraybackslash}p{0.27\textwidth}
			>{\raggedright\arraybackslash}X
			>{\raggedright\arraybackslash}X
		}
		\toprule
		\textbf{Modelling choice} &
		\textbf{Limitations} &
		\textbf{Advantages} \\
		\midrule
		
		Equivalent Euler-Bernoulli/SDOF structural representation &
		No multi-modal, or spatial information &
		Controlled global response with low computational cost \\
		
		\addlinespace
		
		Linear structural response formulation &
		Exclusion of structural response nonlinearities &
		Isolation of EOVs, damage, and sensor-related effects\\
		
		\addlinespace
		
		Sustained/intermittent live load and stochastic band-limited ambient excitation &
		Spatial/moving loads and other specific excitation mechanisms not included &
		Controlled short- and long-term OVs and acquisition-to-acquisition variability\\
		
		\addlinespace
		
		Single equivalent structural temperature affecting $E(T)$ &
		No thermal gradients, thermal strains, restraint effects &
		Controlled environmental influence on stiffness through a known temperature covariate\\
		
		\addlinespace
		
		FAST equivalent stiffness changes &
		No damage localisation or explicit local damage mechanics &
		Known onset and severity for detection and feature-sensitivity studies \\
		
		\addlinespace
		
		SLOW damage as progressive changes in equivalent structural properties &
		Spatial deterioration and nonlinear post-limit response excluded &
		Labelled progressive trend for testing long-term damage identification \\
		
		\addlinespace
		
		Seven SF/M classes applied \textit{a posteriori} &
		No detailed sensor-hardware model or mounting degradation &
		Controlled fault identification and robustness studies independently of structural changes \\
		
		\addlinespace
		
		Limited propagation of parametric uncertainty &
		Geometry, materials, and boundary conditions are mainly deterministic &
		Clearly defined ground truth for better reproducibility and interpretability \\
		
		\addlinespace
		
		Paired static and dynamic response data &
		Limited measurement types and locations &
		Supports comparison and fusion of complementary response quantities \\
		
		\bottomrule
	\end{tabularx}
\end{table*}

\section{Benchmark design}
\label{sec3}
\subsection{Geometry, materials and constraints}
The chosen benchmark structure for simulated monitoring data generation is a steel beam IPE400 supporting a concrete slab that does not contribute to the stiffness of the section. The beam was assumed to belong to a structure located in Italy destined for residential use, and to be fixed at each end. Left-end and right-end supports of the beam are indicated as A and B, respectively. Geometric dimensions, material parameters, boundary constraints, and loading conditions were defined to define the baseline configuration. The characteristics of the structure used as an initial reference are shown in Table~\ref{tab:Table2}, in terms of geometry, materials, loads, and boundary conditions.

\begin{figure}[!ht]
	\includegraphics[width = 1\linewidth]{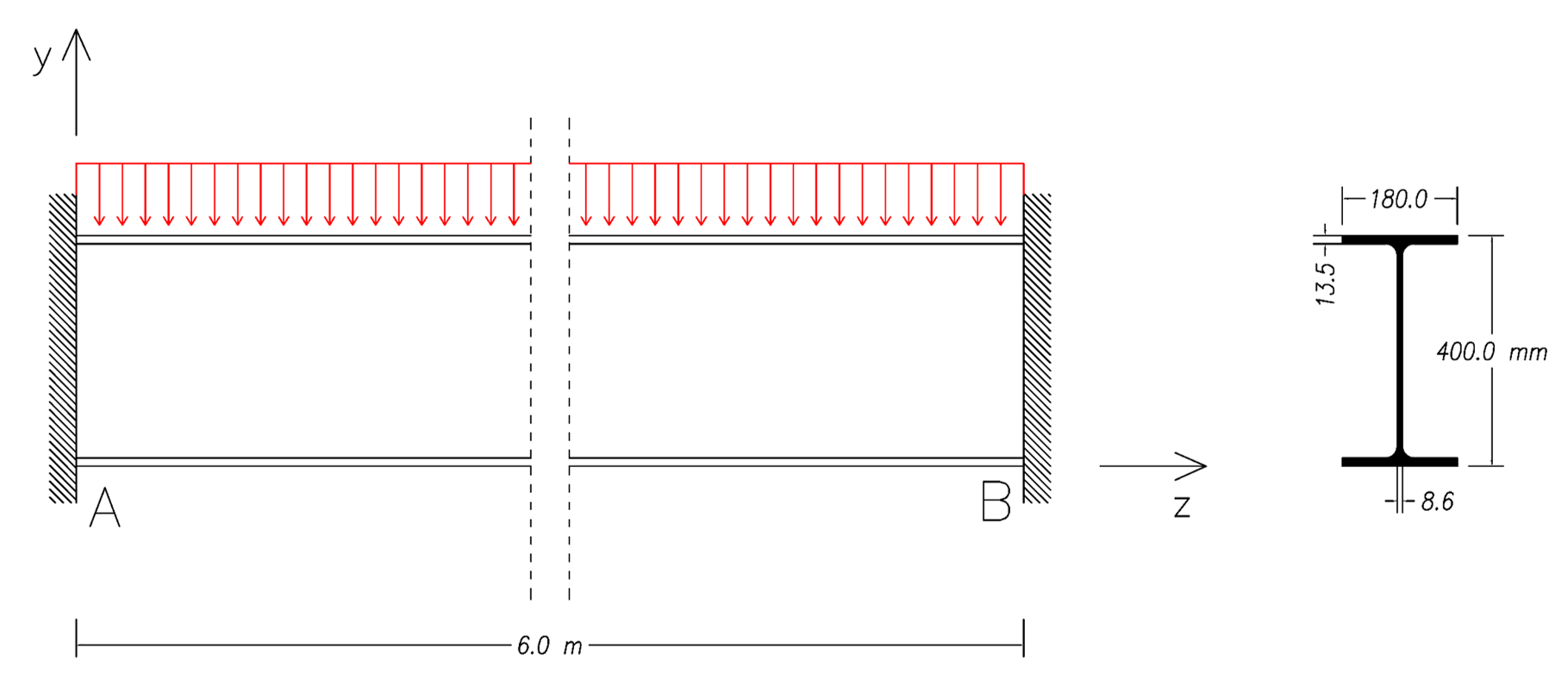}	{\caption{The benchmark steel beam}\label{fig:Figure3bis}}
	\vspace{-22pt}
\end{figure}

\begin{table}[t]
\caption{Geometry, material properties and constraints}
\label{tab:Table2}
\centering
\begin{tabular}{p{0.48\linewidth} c c c}
\toprule
\textbf{Description} & \textbf{Symbol} & \textbf{Quantity} & \textbf{Units} \\
\midrule
\multicolumn{4}{c}{\textbf{Geometry}} \\
\midrule
Length from A to B & $L$ & $6000$ & $\mathrm{mm}$ \\
Beam tributary area & $A_s$ & $30 \cdot 10^{6}$ & $\mathrm{mm}^2$ \\
Thickness of the concrete slab & $t_s$ & $180$ & $\mathrm{mm}$ \\
Cross-section IPE400 & $A$ & $84.46 \cdot 10^{2}$ & $\mathrm{mm}^2$ \\
Section height & $h$ & $400$ & $\mathrm{mm}$ \\
Flange width & $b$ & $180$ & $\mathrm{mm}$ \\
Web thickness & $t_w$ & $8.60$ & $\mathrm{mm}$ \\
Flange thickness & $t_f$ & $13.50$ & $\mathrm{mm}$ \\
Inertia moment around strong axis & $I_{xx}$ & $231.30 \cdot 10^{6}$ & $\mathrm{mm}^4$ \\
Elastic section modulus & $W_{el}$ & $1156 \cdot 10^{3}$ & $\mathrm{mm}^3$ \\
Plastic section modulus & $W_{pl}$ & $1307 \cdot 10^{3}$ & $\mathrm{mm}^3$ \\
\midrule
\multicolumn{4}{c}{\textbf{Materials}} \\
\midrule
Young's modulus steel & $E_0$ & $210000$ & $\mathrm{MPa}$ \\
Yield strength steel S235 & $\sigma_y$ & $235$ & $\mathrm{MPa}$ \\
Steel partial safety coefficient & $\gamma_s$ & $1.05$ & $-$ \\
Yield design strength steel S235 & $f_{yd}$ & $\sigma_y / \gamma_s = 223.9$ & $\mathrm{MPa}$ \\
Density of steel & $d_s$ & $7850$ & $\mathrm{kg}/\mathrm{m}^3$ \\
Density of reinforced concrete & $d_c$ & $2500$ & $\mathrm{kg}/\mathrm{m}^3$ \\

\midrule
\multicolumn{4}{c}{\textbf{Constraints}} \\
\midrule
Horizontal displacement constrained & $x_A, x_B$ & $0$ & $\mathrm{m}$ \\
Vertical displacement constrained & $y_A, y_B$ & $0$ & $\mathrm{m}$ \\
Rotation at the supports constrained & $\phi_A, \phi_B$ & $0$ & $\mathrm{rad}$ \\

\midrule
\multicolumn{4}{c}{\textbf{Reference time-invariant loads}} \\
\midrule
Self-weight beam & $p_b$ & $0.65$ & $\mathrm{kN}/\mathrm{m}$ \\
Self-weight slab & $p_s$ & $22.07$ & $\mathrm{kN}/\mathrm{m}$ \\
Non-structural dead loads & $p_{G2}$ & $6.0$ & $\mathrm{kN}/\mathrm{m}$ \\
Total structural dead load & $p_{G1}$ & $= p_b + p_s = 22.72$ & $\mathrm{kN}/\mathrm{m}$ \\
Total dead load & $p_G$ & $= p_{G1} + p_{G2} = 28.72$ & $\mathrm{kN}/\mathrm{m}$ \\
Reference time-invariant live load & $p_Q$ & $15.0$ & $\mathrm{kN}/\mathrm{m}$ \\

\bottomrule
\end{tabular}
\end{table}

\subsection{Operational conditions}
Changes in the static live load were introduced to simulate fluctuations in the dynamic behaviour induced by variability in normal use as well as exceptional crowding events, and were reflected through variations in the structural mass. The permanent load $p_G$, due to self-weight, was deterministically calculated and kept constant throughout the monitoring period. Conversely, the live load $p_Q$ was not treated as time-invariant (as reported in Table~\ref{tab:Table3}), but was modelled probabilistically according to the Probability Model Code of the Joint Committee on Structural Safety (JCSS) \cite{JCSSProbabilisticModel2021}. Two components of the live load were generated: a sustained (long-term) component $p_{Q,lt}(t)$ and an intermittent (short-term) component $p_{Q,st}(t)$, accounting for occupancy variability and possible crowded events. The total load acting on the structure was defined according to a serviceability limit state (SLS) combination \cite{barocciNormeTecnicheCostruzioni2018}, as per Eq.\ref{eq:eq3}).

\begin{equation}
p_{des}(t) = 1.0\,p_G + 1.0\bigl(p_{Q,lt}(t) + p_{Q,st}(t)\bigr).
\label{eq:eq3}
\end{equation}

The sustained load $p_{Q,lt}(t)$ was considered as a Poisson square wave stationary process, and its intensity, sampled from a Gamma distribution $i_{lt,k} \sim \mathrm{Gamma}(\alpha,\beta)$, remains constant between occupancy changes. The parameters $\alpha$ and $\beta$, shape and rate of the Gamma distribution, respectively, were calculated from the mean and variance as $m_{lt} = \frac{\alpha}{\beta}$,$\sigma_{lt}^2 = \frac{\alpha}{\beta^2}$. Mean $m_{lt}$ and variance $\sigma_{lt}$ were chosen with reference to the equations (2.2) and (2.3) in [8] with parameters from Table 2.2.1 also in \cite{JCSSProbabilisticModel2021} (see Eq.\ref{eq:eq4} and Table\ref{tab:Table3}). The variance was defined as per Eq.\ref{eq:eq4}:

\begin{equation}
\sigma_{\mathrm{lt}} = \sqrt{\sigma_v^2 + \sigma_u^2 \, \kappa \, \frac{A_0}{A}},
\label{eq:eq4}
\end{equation}

where $\kappa$ represents the influence-line shape factor (assumed equal to~1), and $A_0/A$ is the ratio between the reference area and the tributary area $A$ (with $A>A_0 = 20\,\mathrm{m}^2$).

The time intervals between the sustained load changes were calculated by sampling them from an exponential distribution. Mean and variance of the exponential distribution ($m = \frac{1}{\lambda}$ and $\sigma^2 = \frac{1}{\lambda^2}$) used for the rate of change, were calculated from the renewal rate $\lambda$ also provided in Table 2.2 in \cite{JCSSProbabilisticModel2021}. 

The intermitted load $p_{Q,st}(t)$ was instead modelled as a Poisson spike non-stationary process, with intensity also assumed to have a Gamma distribution (with mean $m_{st}$ reported in Table~\ref{tab:Table3} and variance $\sigma_{st}$ from Eq.\ref{eq:eq5} and an exponential distribution for the definition of the time of occurrence. 

\begin{equation}
\sigma_{st} = \sqrt{\sigma_u^2 \, \kappa \, \frac{A_0}{A}}.
\label{eq:eq5}
\end{equation}

\begin{figure}[t]
    \FIG{\includegraphics[width = 1\linewidth]{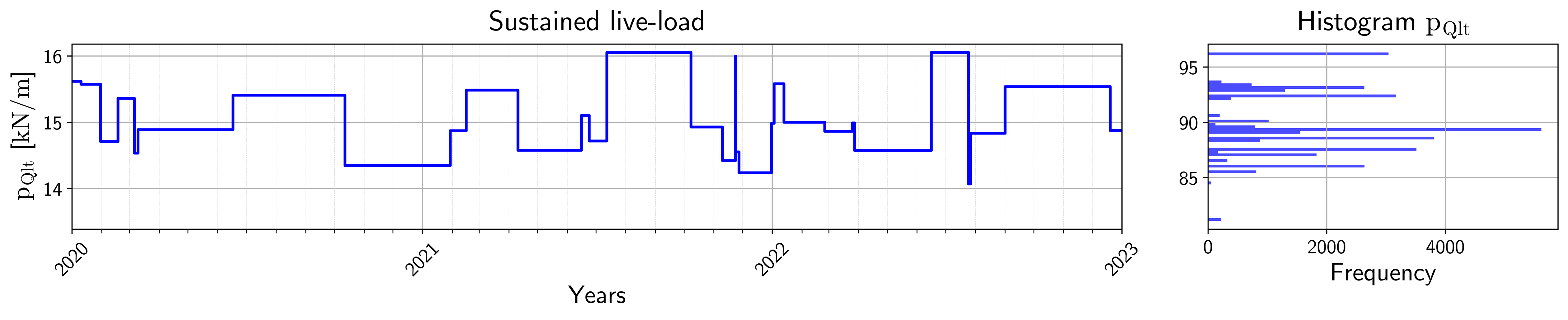}}
    {\caption{Simulated variability of the applied live load in time}\label{fig:Figure4}}
\end{figure}

The scale of the exponential distribution used for the intermitted load was defined as the inverse of the number of occurrences per year ($\nu$). Each intermitted load event duration was considered constant and set to 5 days. The generated load profile is plotted in Fig.\ref{fig:Figure4}. Finally, in order to verify the designed beam considering both an ultimate limit state (ULS) and a serviceability limit state for irreversible effects (SLS) \cite{barocciNormeTecnicheCostruzioni2018} were taken into account (see Table~\ref{tab:Table3}).

The two live load components are combined over the simulation period to represent the total live load \cite{vrouwenvelderJCSSProbabilisticModel2021, costaStochasticLiveLoad2022}. Note that the selected rate of occurrence of the intermitted load is significantly higher than the value recommended in [8], and the mean and rate of change of the sustained load have also been adjusted (strike-through and highlight in Table~\ref{tab:Table4}). These adjustments were made in order to better align with the data generation objective, which is to simulate a realistic operational variability range keeping the ratio between permanent/live load within the calculated reference, while the JCSS guidelines aim at reliability analysis at limit states exceedance.
 
\begin{table}[h]
\caption{Parameter for live loads for used category residential. Some parameters were adjusted with respect to those recommended in~\cite{JCSSProbabilisticModel2021}, reported in brackets}
\centering
\label{tab:Table3}
\centering
\setlength{\tabcolsep}{4pt}
\begin{tabular}{p{0.44\linewidth} c p{0.28\linewidth} c}
\toprule
\textbf{Description} & \textbf{Symbol} & \textbf{Quantity} & \textbf{Units} \\
\midrule
Reference minimum influence area & $A_0$ & $20$ & $\mathrm{m}^2$ \\
Mean load intensity (lt) & $m_{\mathrm{lt}}$ & $3 (0.3)$ & $\mathrm{kN}/\mathrm{m}^2$ \\
Global load intensity std parameter (lt) & $\sigma_v$ & $0.15$ & $\mathrm{kN}/\mathrm{m}^2$ \\
Local load intensity std parameter (lt)  & $\sigma_u$ & $0.3$ & $\mathrm{kN}/\mathrm{m}^2$ \\
Sustained load intensity std parameter & $\sigma_{\mathrm{lt}}$ & $0.07$ & $\mathrm{kN}/\mathrm{m}^2$ \\
Inverse of the renewal rate & $1/\lambda$ & $10 (7)$ & $\mathrm{year}$ \\
Mean load intensity (st) & $m_{\mathrm{st}}$ & $0.3$ & $\mathrm{kN}/\mathrm{m}^2$ \\
Load intensity std parameter (st)
 & $\sigma_U$ & $0.4$ & $\mathrm{kN}/\mathrm{m}^2$ \\
Standard deviation of the load intensity (st) & $\sigma_{\mathrm{st}}$ & $0.03$ & $\mathrm{kN}/\mathrm{m}^2$ \\
Inverse of occurrences per year & $1/\nu$ & $0.2 (1)$ & $\mathrm{year}$ \\
Inter-arrival duration intensity & $d_p$ & $5$ & $\mathrm{days}$ \\
\bottomrule
\end{tabular}
\end{table}

\begin{table}[htbp]
\caption{Extreme values for live loads for the selected profile and serviceability limit checks}
\label{tab:Table4}
\centering
\setlength{\tabcolsep}{4pt}
\begin{tabular}{p{0.44\linewidth} c c c}
\toprule
\textbf{Description} & \textbf{Symbol} & \textbf{Quantity} & \textbf{Units} \\
\midrule
Maximum sustained live load & $p_{Q\mathrm{lt},\max}$ & $16.05$ & $\mathrm{kN}/\mathrm{m}$ \\
Minimum sustained live load & $p_{Q\mathrm{lt},\min}$ & $13.51$ & $\mathrm{kN}/\mathrm{m}$ \\
Maximum intermitted live load & $p_{Q\mathrm{st},\max}$ & $4.18$ & $\mathrm{kN}/\mathrm{m}$ \\
Minimum intermitted live load & $p_{Q\mathrm{st},\min}$ & $0.00$ & $\mathrm{kN}/\mathrm{m}$ \\
Maximum live load & $p_{Q,\max}$ & $\max\!\left(p_Q(t)\right)=19.11$ & $\mathrm{kN}/\mathrm{m}$ \\
Minimum live load & $p_{Q,\min}$ & $\min\!\left(p_Q(t)\right)=11.51$ & $\mathrm{kN}/\mathrm{m}$ \\
Average live load & $p_{Q,\mathrm{avg}}$ & $\mathrm{avg}\!\left(p_Q(t)\right)=15.01$ & $\mathrm{kN}/\mathrm{m}$ \\
Maximum ratio with permanent load & $\dfrac{p_{Q,\max}}{p_G}$ & $67.29$ & $\%$ \\
Minimum ratio with permanent load & $\dfrac{p_{Q,\min}}{p_G}$ & $47.83$ & $\%$ \\
Maximum total load ULS & $p_{\mathrm{des},\mathrm{ULS},\max}$ & $67.20$ & $\mathrm{kN}/\mathrm{m}$ \\
Maximum acting moment ULS & $M_{A,\mathrm{ULS}}$ & $201.60$ & $\mathrm{kN}\,\mathrm{m}$ \\
Maximum deflection due to $p_Q$ & $\delta_{\mathrm{des},Q,\max}$ & $1.32$ & $\mathrm{mm}$ \\
\midrule
Average design load ULS & $p_{\mathrm{des},\mathrm{ULS}}$ & $61.04$ & $\mathrm{kN}/\mathrm{m}$ \\
Average design load SLS & $p_{\mathrm{des},\mathrm{SLS}}$ & $43.72$ & $\mathrm{kN}/\mathrm{m}$ \\
Resisting moment & $M_R$ & $W_{\mathrm{el}}\cdot f_{yd}=258.84$ & $\mathrm{kN}\,\mathrm{m}$ \\
Acting moment ULS & $M_{A,\mathrm{ULS}}$ & $\dfrac{p_{\mathrm{des},\mathrm{ULS}}\,L^2}{12}=183.12$ & $\mathrm{kN}\,\mathrm{m}$ \\
\midrule
Deflection limit for $p_G$ & $\delta_{\mathrm{lim},G}$ & $\dfrac{L}{250}=24$ & $\mathrm{mm}$ \\
Deflection limit for $p_Q$ & $\delta_{\mathrm{lim},Q}$ & $\dfrac{L}{300}=20$ & $\mathrm{mm}$ \\
Maximum deflection due to $p_G$ & $\delta_{\mathrm{des},G}$ & $\dfrac{p_G L^4}{384\,E\,I_x}=1.99$ & $\mathrm{mm}$ \\
Maximum deflection due to $p_Q$ & $\delta_{\mathrm{des},Q}$ & $\dfrac{p_Q L^4}{384\,E\,I_x}=1.04$ & $\mathrm{mm}$ \\
\bottomrule
\end{tabular}
\end{table}
\vspace{-10pt}

\subsection{Environmental conditions}
The Young’s Modulus $E$ of steel was varied in each simulation as per Eq.\ref{eq:eq6}), in which $E_0$ is the reference value of $E$ at $20^\circ\mathrm{C}$, $T^{*}$ the prescribed equivalent structural temperature, $\alpha_T$ an amplification parameter added to scale the temperature effect. The constants $e_1$, $e_2$, $e_3$, and $e_4$ were experimentally obtained in \cite{seifTemperatureDependentMaterialModeling2016}.

\begin{equation}
E(T) = E_0 \left( 1 - \alpha_T \cdot T^{*} \right)
\exp\left[
-\frac{1}{2}\left(\frac{T^{*}}{e_3}\right)^{e_1}
-\frac{1}{2}\left(\frac{T^{*}}{e_4}\right)^{e_2}
\right]
\label{eq:eq6}
\end{equation}

\begin{figure}[htbp]
    \FIG{\includegraphics[width = 1\linewidth]{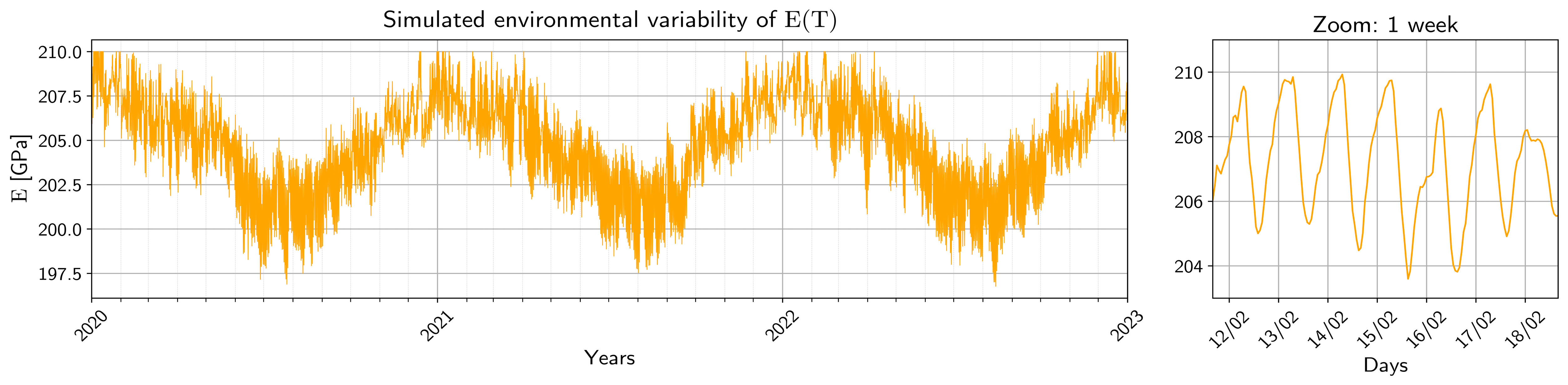}}
    {\caption{Simulated variability of the steel Young’s Modulus with temperature}\label{fig:Figure5}}
\end{figure}

The time-series of temperature was obtained from the recording of a sensor in the vicinity of the Structure and material testing Lab of the University of Florence, the recorded temperature series was used as a proxy for a spatially uniform structural temperature. To accentuate the effect of temperature, an amplification parameter ($\alpha_T = 0.0015$ in Eq.\ref{eq:eq6}) was included to increase the effect of environmental variability on Young's modulus (see Fig.\ref{fig:Figure5}). The recorded temperature series was used as a proxy for a spatially uniform structural temperature. The applied amplification was used to obtain a variability in natural frequency comparable in magnitude to that reported in the literature and observed in monitored structures. The resulting frequency variation and temperature–frequency correlation were calculated and they fall within the ranges reported for real monitoring cases, although both quantities vary considerably among structures and individual modes and may become strongly nonlinear under specific environmental conditions. The comparison was not intended to imply that the simplified temperature model reproduces the complete thermo-mechanical behaviour of a real structure, but rather to support the adopted magnitude and direction of the environmental variability as a reasonable controlled benchmark condition. In fact, although the reference beam has fixed end conditions, thermal expansion, thermally induced axial forces, and their contribution to geometric stiffness are not included. The simulated temperature dependence therefore represents the imposed material-stiffness effect alone, rather than the complete thermal response of an axially restrained beam.

\subsection{Dynamic and static response}
\begin{figure}[htbp]
    \FIG{\includegraphics[width = 1\linewidth]{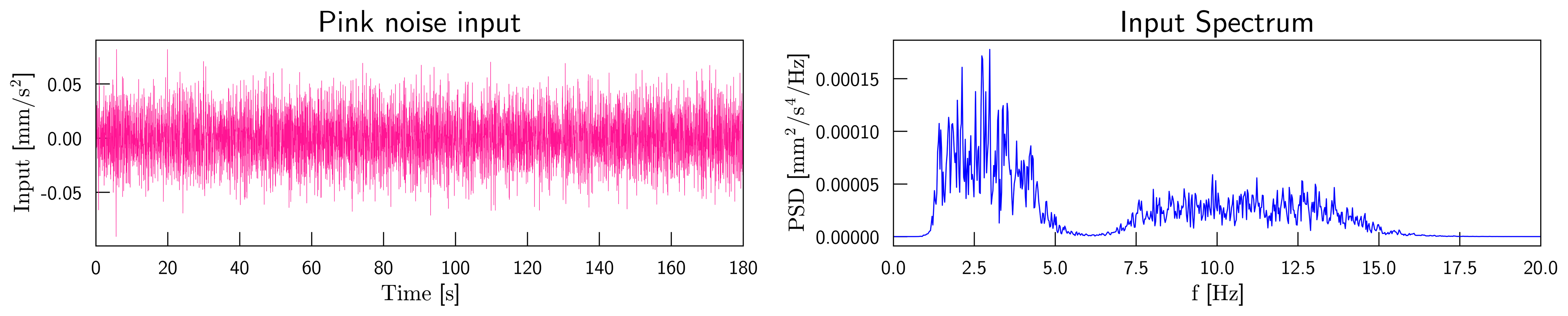}}
    {\caption{Example of coloured input for ambient vibration simulation}\label{fig:Figure6}}
\end{figure}

For dynamic response simulation, specific frequency bands were targeted by two band-pass filters to simulate the effects of human activities and vehicular traffic. To differentiate the two actions, two Gaussian inputs were generated for each hour of monitoring (acquisition), their standard deviation was randomly sampled within specific ranges (see Table\ref{tab:Table5})), filtered and then superimposed to get the $\omega_{AV}$ input. An example is provided in Fig.\ref{fig:Figure6}).

\begin{table}[htbp]
\caption{Ambient vibration simulation}
\label{tab:Table5}
\centering
\setlength{\tabcolsep}{6pt}
\renewcommand{\arraystretch}{1.2}

\begin{tabular}{p{0.40\linewidth} c c c}
\toprule
\textbf{Description} & \textbf{Symbol} & \textbf{Quantity} & \textbf{Units} \\
\midrule
Sampling frequency & $f_s$ & $100$ & $\mathrm{Hz}$ \\
Signal length & $l_s$ & $180$ & $\mathrm{s}$ \\
Time interval & $dt$ & $0.01$ & $\mathrm{s}$ \\
Frequency resolution & $\Delta f$ & $0.0056$ & $\mathrm{Hz}$ \\
\midrule
\multirow{2}{*}{Human interaction}
& $f_{\mathrm{range}}$ & $1.2 \;-\; 4.8$ & $\mathrm{Hz}$ \\
& $\sigma_{\mathrm{range}}$ & $5\times10^{-6} \;-\; 5\times10^{-4}$ & $\mathrm{m}/\mathrm{s}^2$ \\
\midrule
\multirow{2}{*}{Traffic}
& $f_{\mathrm{range}}$ & $7 \;-\; 15$ & $\mathrm{Hz}$ \\
& $\sigma_{\mathrm{range}}$ & $1\times10^{-6} \;-\; 2\times10^{-4}$ & $\mathrm{m}/\mathrm{s}^2$ \\
\bottomrule
\end{tabular}
\end{table}

A state-space approach was used to solve the equation of motion to calculate the dynamic structural response of the SDOF system, given the calculated equivalent stiffness dependent on $E$ and therefore dependent on the temperature, $k(T(t))$, the equivalent mass calculated from the time-varying imposed static load, $m(l(t))$, the damping coefficient $c$ ($c = 2 \zeta m \sqrt{k/m}$ with $\zeta = 5\%$), and the defined input vibration $\omega_{\mathrm{AV}}$. The method allows the addition of measurement noise $\omega_M$, even if in the current data generation it was not introduced. The state variables, displacement $x_1 = x(t)$ and velocity $x_2 = \dot{x}(t)$, were substituted into the equation of motion, leading to a system of two first-order differential equations (see Eq.~\ref{eq:eq7}).

\begin{equation}
\dot{\mathbf{x}}(t) =
\begin{bmatrix}
0 & 1 \\
-\dfrac{k}{m} & -\dfrac{c}{m}
\end{bmatrix}
\begin{bmatrix}
x_1 \\
x_2
\end{bmatrix}
+
\begin{bmatrix}
0 \\
\dfrac{1}{m}
\end{bmatrix}
(\omega_{\mathrm{AV}}\, m)
+
\begin{bmatrix}
0 \\
\dfrac{1}{m}
\end{bmatrix}
\omega_M
\label{eq:eq7}
\end{equation}

It was verified that the maximum weighted acceleration generated by the simulation was within the limits for human comfort for both day and night vibrations, according to UNI 9614--2017~\cite{UNI961420172017} and ISO 2631--2014~\cite{UNIISO263112014}.

\begin{figure}[htbp]
    \FIG{\includegraphics[width = 1\linewidth]{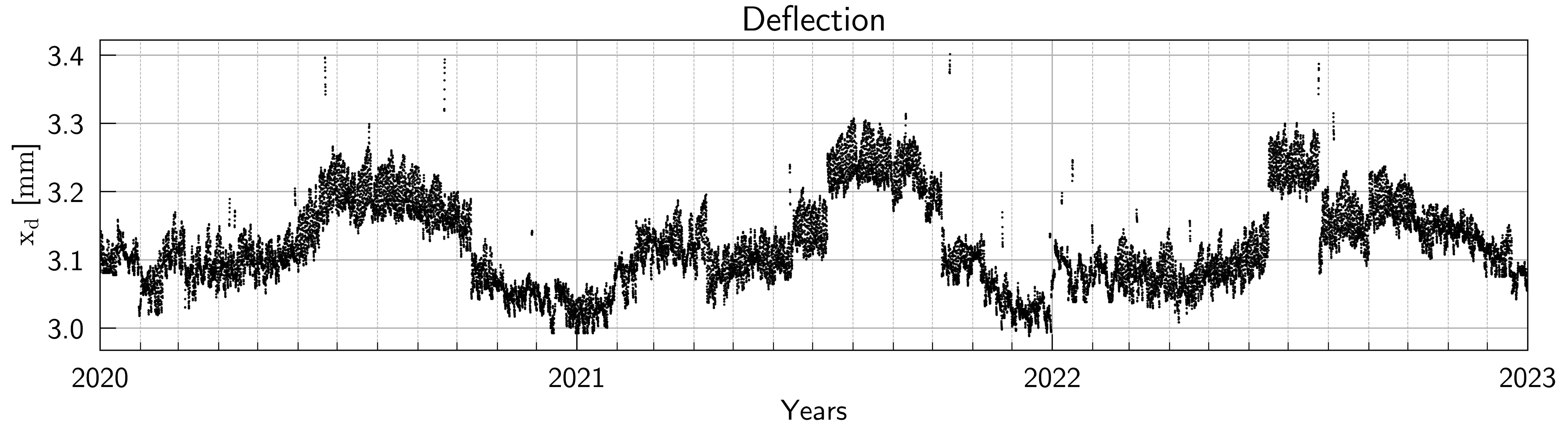}}
    {\caption{Simulated static monitoring: mid-span deflection with the effect of EOVs}\label{fig:Figure7}}
\end{figure}

For the static response simulation, the single value of the deflection at mid-span for the reference fixed-fixed Euler-Bernoulli beam under the time-varying uniformly distributed load was calculated for the static monitoring simulation. 
The formula used for the serviceability limit checks was also used for the generation of the static response (see Table~\ref{tab:Table4}). 
The deflection results for the undamaged condition simulated for the 3 years of monitoring are shown in Fig.~\ref{fig:Figure7}.

\subsection{Damage simulation}
Damage can be considered as a phenomenon which leads to a variation or a deviation from the initial or undamaged configuration, affecting the properties of the structural system ($k_i, m_i$ in the present case). Damage was represented through reductions in equivalent structural stiffness, see  Figure~\ref{fig:Figure2}, these reductions were imposed without explicitly modelling a damage location or the physical mechanism responsible for the change. Each acquisition therefore represents the low-amplitude response of the structure at a given structural state, with the corresponding equivalent properties kept constant during the acquisition and updated between successive monitoring states. The formulation is intended to characterise the response after a change in structural condition has occurred. It was considered affecting only $I$ with a decay rate of $r = \Delta I / I_{\mathrm{UD}}$, yielding a stiffness variation from the undamaged (UD) condition equal to $\Delta k = k_{\mathrm{UD}}\, r$. Since the natural frequency depends on stiffness as $f\propto\sqrt{k}$, the corresponding frequency variation followed as $\frac{\Delta f}{f_{\mathrm{UD}}} = \sqrt{r}$. 
To contextualize the imposed stiffness reductions within the mechanical capacity of the reference section, a reference capacity check was used, by comparing the elastic resisting moment $M_{R,\mathrm{el}}$ with the maximum acting bending moment $\max(M_A(t))$. Under the additional assumption that elastic moment resistance decreases in proportion to the imposed effective inertia reduction, Eq.~\ref{eq:eq8} defines the reduction at which the reference acting moment equals the reduced elastic resistance, considered as a critical decay rate. 

\begin{equation}
r_{\mathrm{pl}} = 
\frac{M_{R,\mathrm{el}} - \max(M_A(t))}{M_{R,\mathrm{el}}}.
\label{eq:eq8}
\end{equation}

 Leading to the to corresponding stiffness and frequency limits 
$\Delta k_{\mathrm{pl}} = k_{\mathrm{UD}} r_{\mathrm{pl}}$ and 
$\Delta f_{\mathrm{pl}} = f_{\mathrm{UD}} \sqrt{r_{\mathrm{pl}}}$. Similarly, serviceability limits under permanent and live loads were derived assuming the deflection scaled inversely with stiffness ($\delta \propto 1/k$). For each deflection limit $\delta_{\mathrm{lim}}$, the associated decay rate is as per Eq.~\ref{eq:eq9}.

\begin{equation}
r_{\mathrm{lim}} = 1 - \frac{\delta_{\mathrm{des}}}{\delta_{\mathrm{lim}}},
\label{eq:eq9}
\end{equation}

\begin{figure}[htbp]
    \FIG{\includegraphics[width = 1\linewidth]{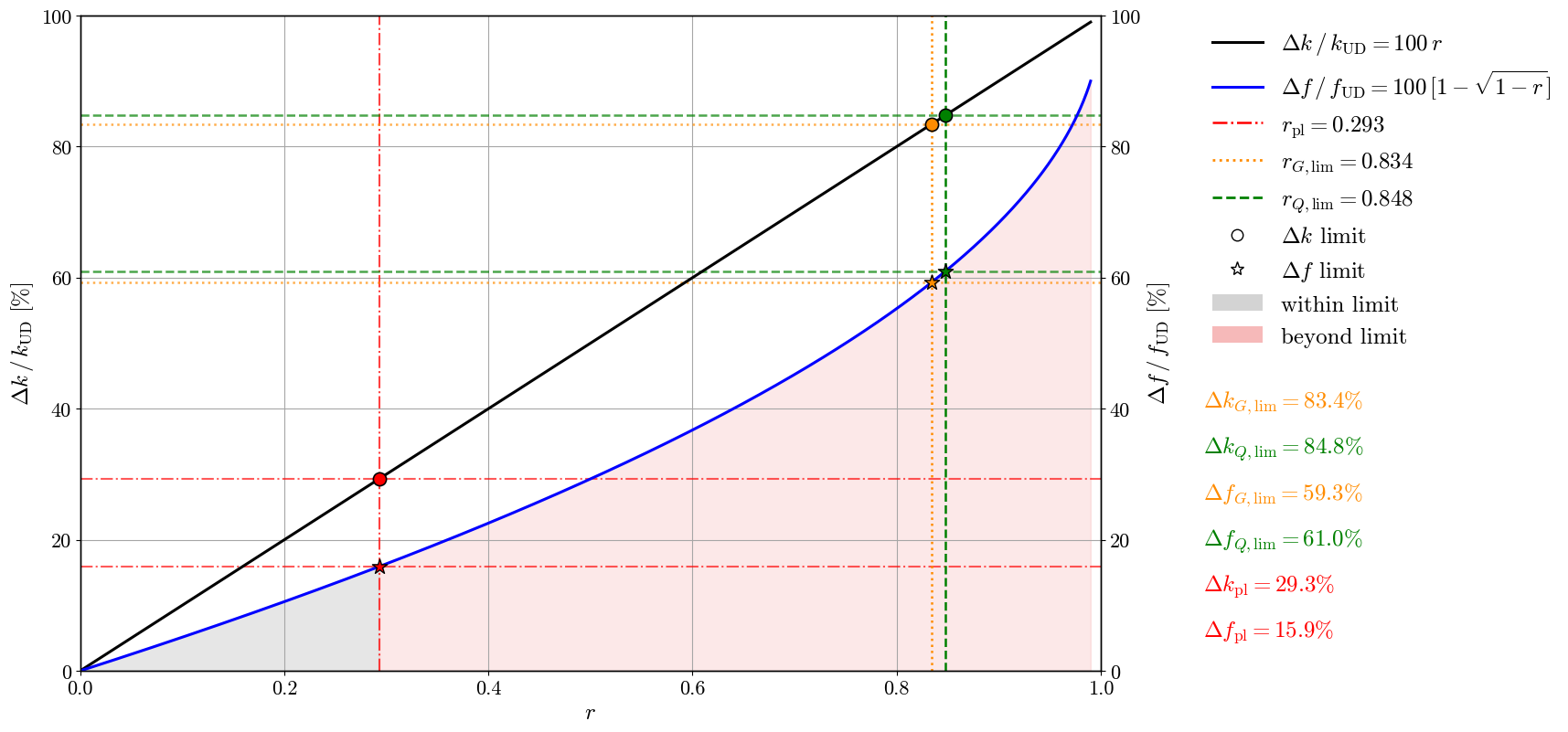}}
    {\caption{Variation of stiffness $(\Delta k/k_{\mathrm{UD}})$ and frequency $(\Delta f/f_{\mathrm{UD}})$ as a function of the rate of decay of the stiffness r}\label{fig:Figure8}}
\end{figure}

 From Eq.~\ref{eq:eq9} the corresponding stiffness and frequency variations follow directly as $\Delta k_{\mathrm{lim}}/k_{\mathrm{UD}} = r_{\mathrm{lim}}$ and 
$\Delta f_{\mathrm{lim}}/f_{\mathrm{UD}} = \sqrt{r_{\mathrm{lim}}}$. These limits are computed and summarized in Figure~\ref{fig:Figure8}. The reference capacity-based stiffness-reduction threshold was found to be $\Delta k_{\mathrm{pl}} = 29.3\%$, while the serviceability limits in terms of stiffness and frequency variations were well beyond the plastic threshold.

Under the additional assumption that elastic moment resistance decreases in proportion to the imposed effective inertia reduction, Eq. (8) defines the reduction at which the reference acting moment equals the reduced elastic resistance. This is a contextual threshold, not a simulated transition to plastic response.

\subsubsection{Fast-varying damage}
 Considering the modelling approach described, three configurations of sudden, FAST damage were simulated, as shown in Table~\ref{fig:Figure8} in Section~\ref{sec:datades} describing the generated data. A $1\%$ reduction in $k_i$ corresponded approximately to a $0.5\%$ decrease in the analytical natural frequency $f_n$ and to a $1\%$ change in the midspan resisting moment $M_R$, assuming stiffness loss modelled as inertia reduction. With reference to Figure~\ref{fig:Figure8}, the listed percentage reductions of stiffness were chosen restricting the variability of interest to the elastic range. Since EOVs were found to induce frequency variations of up to about $4\%$ and $6\%$, respectively, and FAST-induced stiffness drops was simulated to cumulate with these effects, they were generally kept below $28\%$, with only a limited portion of one dataset reaching $\Delta k = 32\% > \Delta k_{\mathrm{pl}}$. This choice was made to allow the evaluation of the detectability of damage as it approaches, and eventually exceeds, the plastic threshold.  All responses were nevertheless generated using the linear model. The severe scenario therefore supports testing the detectability of large imposed property changes, rather than the identification of yielding or post-yield behaviour.

\subsubsection{Slow-varying damage}
 Damage was also modelled as diffused and slow-varying, representing corrosion as a progressive and uniform thickness reduction of the cross-section at a prescribed rate. This gradual section loss meant a decrease in inertia and, consequently, stiffness, as well as a smaller reduction in mass due to volume loss. SLOW damage therefore translates into a progressive reduction of the equivalent stiffness $k_i$ and, to a lesser extent, of the mass $m_i$ across acquisitions, reflecting the gradual deterioration of structural properties over time.

A corrosion-based section-loss model was selected to generate progressive changes in equivalent structural properties. Its evolution is non-linear: corrosion progresses rapidly at early stages due to direct exposure to environmental agents, then slows as a protective oxide layer forms and partially shields the surface. Over longer periods, the extent of corrosion depends on the penetration of water and reactive substances through defects in this layer. To account for environmental influence, a relationship proposed in 2007~\cite{klinesmithEffectEnvironmentalConditions2007} was adopted, as it explicitly models the interaction between corrosion and environmental variables (see Eq.~\ref{eq:eq10}).

\begin{equation}
d(t) = A \cdot t^{B} \cdot \left( \frac{\mathrm{TOW}}{C} \right)^{D}
\cdot \left( 1 + \frac{[\mathrm{SO}_2]}{E} \right)^{F}
\cdot \left( 1 + \frac{[\mathrm{Cl}]}{G} \right)^{H}
\cdot e^{J(T + T_0)}
\label{eq:eq10}
\end{equation}

The cumulative corrosion loss is expressed in Eq.~\ref{eq:eq10}., in which:
\begin{itemize}
\item $d$ is the corrosion loss expressed in $[\mu\mathrm{m}]$ as a function of $t$, the exposure time $[\mathrm{years}]$, considering here the exposure to start after 1.5 years from the beginning of monitoring;
\item $\mathrm{TOW}$ is the annual time of wetness $[\mathrm{h}/\mathrm{year}]$, the fraction of time that the metal surface remains wet. For simplicity, the TOW was approximated as the overall time in which the RH is greater than $80\%$ and the temperature is greater than $0^{\circ}\mathrm{C}$~\cite{disarnoEffectAtmosphericCorrosion2021} within one year;
\item $\mathrm{SO}_2$ and $\mathrm{Cl}$ are respectively the sulphur dioxide concentration and the chloride deposition rate for non-marine environments;
\item $T_{\mathrm{avg}}$ is the average temperature in $[^{\circ}\mathrm{C}]$ within one year;
\item $A,B,D,E,F,G,H,J$ and $T_0$ are coefficients proposed in~\cite{klinesmithEffectEnvironmentalConditions2007} adapted for the present case.
\end{itemize}

\begin{figure}[!ht]
	\FIG{\includegraphics[width = 1\linewidth]{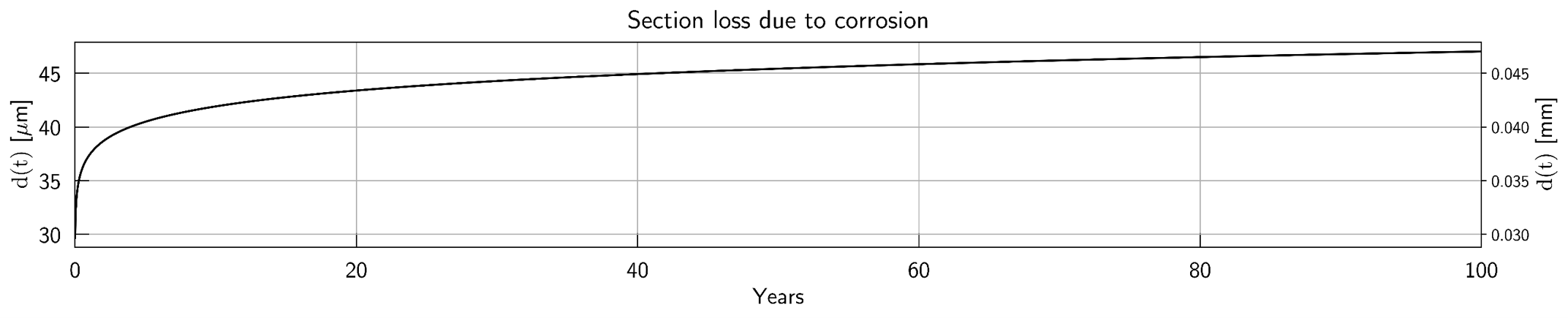}}
	{\caption{Corrosion model comparison for corrosivity level ISO9223 C3 Medium: Temperate zone with medium pollution ~\cite{disarnoEffectAtmosphericCorrosion2021}}\label{fig:Figure8bis}}
	\vspace{-22pt}
\end{figure}

The corrosivity category adopted was “ISO9223 C3 Medium: Temperate zone with medium pollution” $(5 < \mathrm{SO}_2,[\mu\mathrm{g}/\mathrm{m}^3] < 30)$ with low chloride effect, representative of urban or mildly coastal environments. Although the selected model may overestimate long-term corrosion, it provided a relationship explicitly linked to environmental parameters. Using the average environmental values for 2020, the corrosion rate was estimated as $47.03\,\mu\mathrm{m}/\mathrm{year}$ ($0.04703\,\mathrm{mm}/\mathrm{year}$), 
corresponding to approximately $0.2\,\mathrm{mm}$ over the 1.5 year of "damaged" monitoring period and to a $35\%$ flange thickness loss over a hypothetical 100-year exposure. Three corrosion scenarios were analysed: the estimated rate and accelerated rates of $10\times$ and $20\times$, including their effects on mass and stiffness. The resisting moment at the end of the monitoring period was also evaluated: the structure remains in the elastic range for the first level, reaches the plastic range for the second, and does not satisfy the safety check for the third. 
The corrosion model produces a nonlinear evolution of the section loss over the monitoring period (see Figure~\ref{fig:Figure8bis}). For each resulting damage scenario, however, the response to low-amplitude ambient vibration is still simulated through the corresponding updated mass and stiffness within the linear Euler-Bernoulli/SDOF formulation. Therefore, even if the most severe corrosion scenario exceeds the adopted safety threshold, it should be interpreted as an extrapolated condition with progressively reduced equivalent structural properties, rather than as a simulation of the nonlinear response of the structure approaching failure.

\subsection{Sensor faults and malfunctions simulation}
\begin{figure}[!ht]
    \FIG{\includegraphics[width = 1\linewidth]{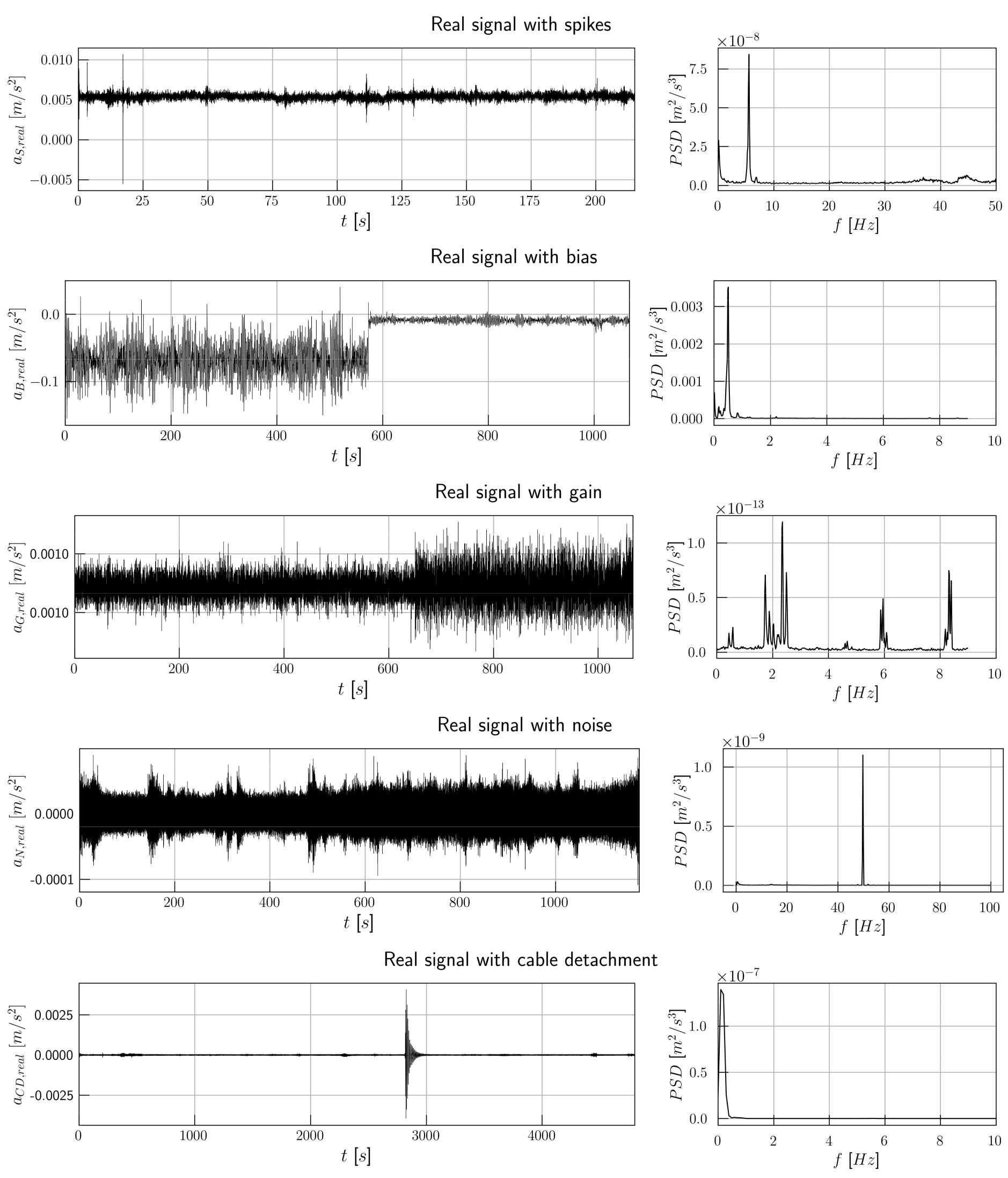}}
    {\caption{Samples for real data acquisitions for the simulated sensor faults and/or malfunctions}\label{fig:Figure9}}
\end{figure}

\begin{figure}[!h]
    \FIG{\includegraphics[width = 1\linewidth]{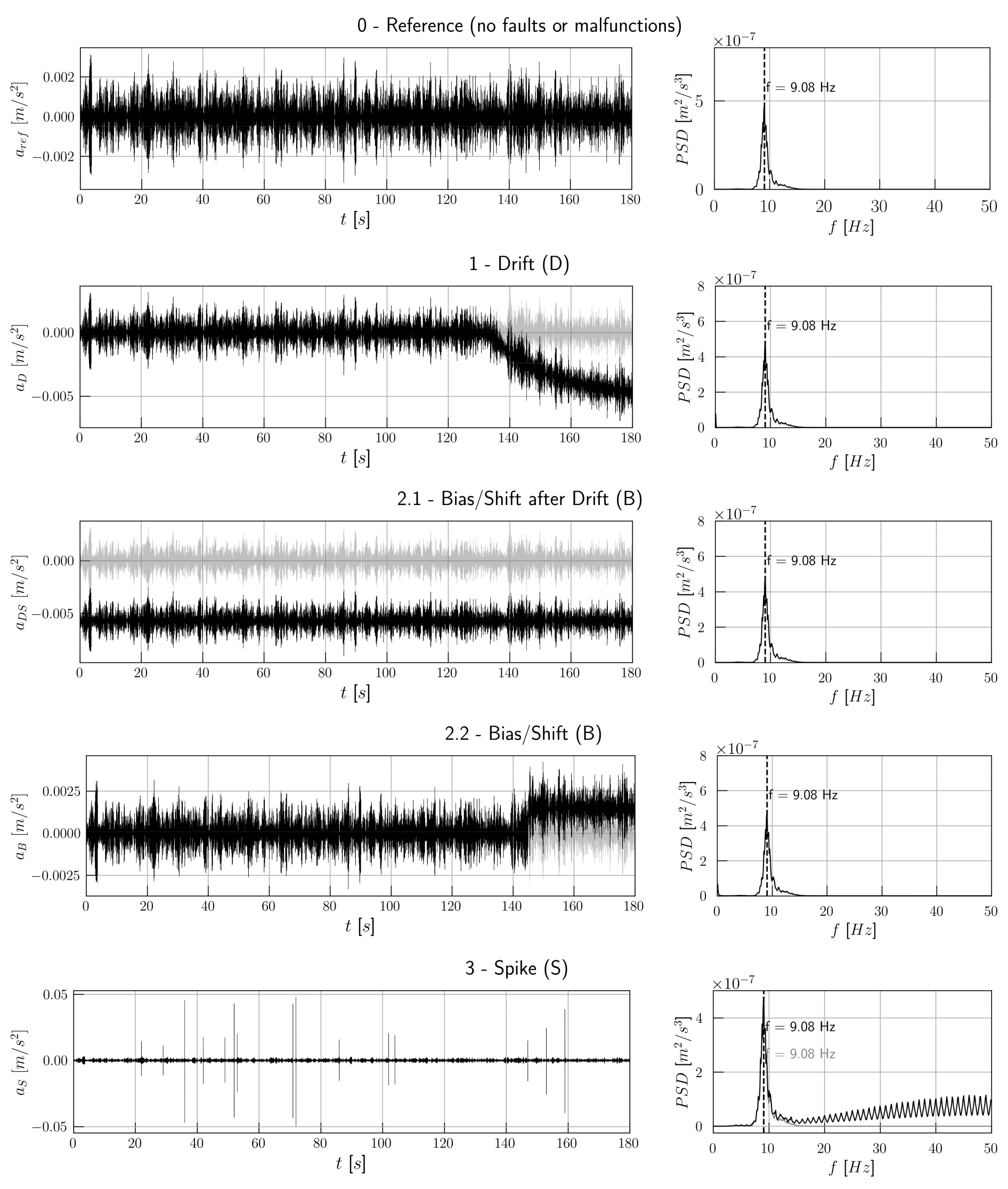}}
    {\caption{Example of generated SF/M in order from top to bottom: (0) reference acceleration signal, no faults or malfunctions (1) drift, (2) bias, (3) spikes. Each image shows the output in the time domain and the frequency domain}\label{fig:Figure10}}
\end{figure}

\begin{figure}[!h]
    \FIG{\includegraphics[width = 1\linewidth]{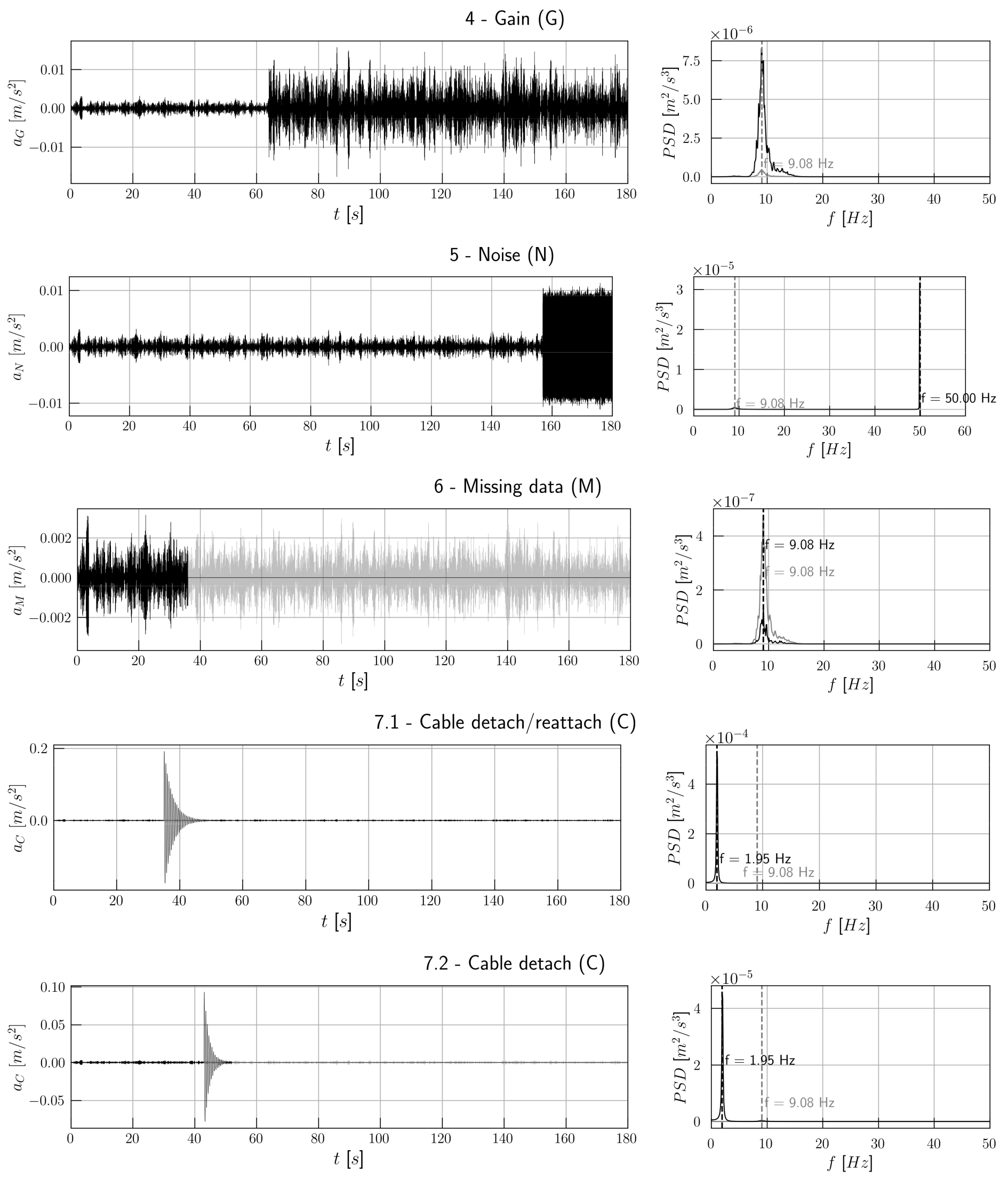}}
    {\caption{Example of generated SF/M in order from top to bottom: (4) gain, (5) noise, (6) missing data, and temporary (7.1) or permanent (7.2) cable detachment. Each image shows the output in the time domain and the frequency domain}\label{fig:Figure11}}
\end{figure}

Sensor faults or malfunctions were simulated by introducing typical anomalous patterns into the acceleration signals. Similar patterns could be used to contaminate also the displacement measurements across the whole time series, with the exception of a few faults or malfunctions that are specific to accelerometers (e.g.\ cable detachment). Seven signal-anomaly categories were selected from the cited literature ~\cite{javaidMachineLearningAlgorithms2019, ahmerFailureModeClassification2022, oncescuSelfsupervisedClassificationAlgorithm2023} and on the authors’ experience with real monitoring data (see Figure~\ref{fig:Figure9}). An illustration of all simulated sensor faults and malfunctions is displayed in Figures~\ref{fig:Figure10} and ~\ref{fig:Figure11}, distinguishing seven categories of sensor faults/malfunctions (Table~\ref{tab:Table6}):
\begin{enumerate}
    \item {Drifting (D):} gradual change in the sensor's output over time;
    \item {Bias or Shift (B):} consistent offset in the sensor output, they can occur also as a result of drift in previous acquisitions;
    \item {Spikes (S):} large amplitude value at a specific instant;
    \item {Gain (G):} amplification or attenuation of the signal;
    \item {Noise (N):} overlaid Gaussian noise onto the signal;
    \item {Missing Data (M):} Gaps within the time series;
    \item {Cable Detachment (C):} sinusoidal wave that rapidly decreases in amplitude, simulating a sensor that disconnects either momentarily or completely.
\end{enumerate}

\begin{table}[htbp]
\caption{Typologies of sensor faults simulated}
\label{tab:Table6}
\centering
\setlength{\tabcolsep}{4pt}
\renewcommand{\arraystretch}{1.15}
\begin{tabular}{p{0.04\linewidth} p{0.38\linewidth} p{0.15\linewidth} p{0.30\linewidth}}
\toprule
\textbf{Class} & \textbf{Formulation} & & \textbf{Ranges of variability} \\
\midrule

D &
$a_D(\tau) = a(\tau) + s_D \cdot \exp\!\left(-\dfrac{\tau}{l_s/10}\right) + s_D$ &
$\tau \in [\tau_{st}, \tau_{end}]$ &
$s_D = 1e^{-3} \div 2e^{-3}\,[\mathrm{m}/\mathrm{s}^2]$ \\

B &
$a_B(\tau) = a(\tau) + s_B$ &
$\tau \in [\tau_{st}, \tau_{end}]$ &
$s_B = 1.2 \div 1.7\,[\mathrm{m}/\mathrm{s}^2]$ \\

S &
$a_S(\tau) = a(\tau) + s_S \cdot (\Delta \tau_S)$ &
$\tau \in [\tau_{st}, \tau_{end}]$ &
$\begin{aligned}
& s_S = 5e^{-2} \div 5e^{-3}\,[\mathrm{m}/\mathrm{s}^2] \\
&\Delta \tau_S = 0.1 f_s \div 0.2 f_s\;[-]
\end{aligned}$ \\

G &
$a_G(\tau) = a(\tau) \cdot s_G$ &
$\tau \in [\tau_{st}, \tau_{end}]$ &
$s_G = 1.1 \div 3.1\,[\mathrm{m}/\mathrm{s}^2]$ \\

N &
$a_N(\tau) = a(\tau) + \eta(\tau)$ &
$\tau \in [\tau_{st}, \tau_{end}]$ &
$s_N = 1e^{-3} \div 1e^{-5}\,[\mathrm{m}/\mathrm{s}^2]$ \\

M &
$a_M(\tau) = \text{null or missing}$ &
$\tau \in [\tau_{st}, \tau_{end}]$ &
$\tau_{st} = 0.8 \div 0.2\,\Delta t\;[-]$ \\

CD &
$a_C(\tau) = s_C \sin(2\pi f_C \tau)\,\exp(-\lambda_C \tau)$ &
$\tau \in [\tau_{st}, \Delta \tau_C]$ &
$\begin{aligned}
f_C &= 0.5 \div 1.5\,[\mathrm{Hz}] \\
s_C &= 0.05 \div 0.2\,[\mathrm{m}/\mathrm{s}^2] \\
\lambda_C &= 0.02 \div 0.05\,[1/\mathrm{s}]
\end{aligned}$ \\

\bottomrule
\end{tabular}
\end{table}

Some of the chosen SF/M categories may apply to any vibration sensor, regardless of its type, while others are specific to certain typologies of sensors (e.g.\ cable detachment does not apply to wireless sensors). The categories were chosen based on their recurrence in the real data observed, but also to obtain a set of classes impacting damage sensitive features differently, as they show different effects in the frequency and time domain. There are additional acquisition-related effects that are not represented, for example changes in the coupling between the sensor and the structure due to degradation, loosening, or temperature dependence of the sensor mounting condition ~\cite{hummel2025}. Such effects may alter the measured amplitude and frequency response ~\cite{randall2020} and could be incorporated as an additional SF/M class in future extensions.

 For each type, the mathematical formula for the simulation is provided in Table~\ref{tab:Table6}. In the formulas, $a(\tau)$ denotes the original signal, while $a_{*}(\tau)$ indicates the modified signal post-fault application, with variations dependent on the time progression $\tau$ within each simulated acceleration acquisition, different from the time progression within the monitoring period $t$ mentioned beforehand. The scale factor $s_{*}$ quantifies the extent of each malfunction, calibrated according to the magnitude of measured accelerations. $L_s$ represents the length of the signal, $t_{\mathrm{st}}$ marks the starting point for fault or malfunction application, randomly chosen, and $t_{\mathrm{end}}$ signifies the end of the signal time history. Specific durations for spikes and cable detachment are denoted by $\Delta t_{*}$. Gaussian noise $\eta(\tau)$ is characterized by a zero mean and a standard deviation $s_N$ that mirrors the sensor noise level. In cable detachment simulations, $s_C$ defines the initial amplitude of the sinusoidal wave, $f_C$ its frequency, and $\lambda_C$ its decay rate, which leads to a rapid amplitude decrease following detachment onset. Spikes were simulated as single acquisition faults, as well as instances of cable detachment and immediate reattachment. All the other faults were simulated as prolonged effects over a randomly selected range of acquisition between 5 and 25. 

\section{Data generation}
\label{sec:datagen}
Leveraging the capabilities of parallel computing, the benchmark utilized the processing power of an Intel Core i7-8750H processor with 12 cores at 2.2\,GHz to simulate the three years of monitoring data in 2.5 hours, utilizing Python~3.11 and the \texttt{Ray} package for implementation. A comprehensive summary of the data generation process with parallel computing is detailed in algorithm\ref{alg:data_generation}. All typologies of the generated sub-dataset are listed in Table~\ref{tab:generated_datasets}. Minimal fluctuations of the generated input spectrum could cause variabilities in the output spectrum and influence the subsequent feature extraction process such as natural frequency identification. In order to attenuate this effect and keep the inherent error within consistent numerical limits, each dynamic simulation was repeated up to ten times, with a tolerance threshold established at $1\%$ of the reference frequency for the simulations. This value is consistent with realistic fluctuations of the extracted frequencies. If the simulations did not achieve a frequency within this margin, the generated signal providing the closest frequency value among the ten iterations was selected. The algorithm used for data generation is elaborated in Algorithm~\ref{alg:data_generation} below.

\begin{algorithm}[htbp]
\caption{Dynamic Data Generation with Parallel State-Space Simulations}
\label{alg:data_generation}

\begin{algorithmic}[1]

\State Load $n_{\mathrm{CPUs}}$ samples from matrix $M \in \mathbb{R}^{n_a \times 2}$
\State Generate time series with hourly intervals and filter leap years
\State Setup methods for input generation, state-space simulation, frequency check

\State Initialize parallel computation on $n_{\mathrm{CPUs}}$

\For{each core $c = (1,\dots,n_{\mathrm{CPUs}})$}
    \State Select one sample $(k_i, m_i)$
    
    \While{$n_{\mathrm{sim}} < n_{\mathrm{sim,max}} = 10$}
    
        \State Generate input signal with length $n_s$
        
        \State Apply passband filter (human interaction, traffic)
        
        \State Solve state-space equations for $(k_i, m_i)$
        
        \State Compute Welch PSD 
        
        \State Extract dominant frequency (peak PSD)
        
        \If{$\dfrac{f_{n,\mathrm{extr}}}{f_{n,\mathrm{anal}}} < 1\%$}
            \State Select acceleration and break
        \Else
            \State Increment $n_{\mathrm{sim}}$
        \EndIf
        
    \EndWhile
    
    \State Select acceleration corresponding to best frequency    
    \State Write selected acceleration to HDF5 file
    
\EndFor

\State End parallel computing

\end{algorithmic}
\end{algorithm}

\section{Dataset description}
\label{sec:datades}
\begin{figure}[!h]
    \FIG{\includegraphics[width = 1\linewidth]{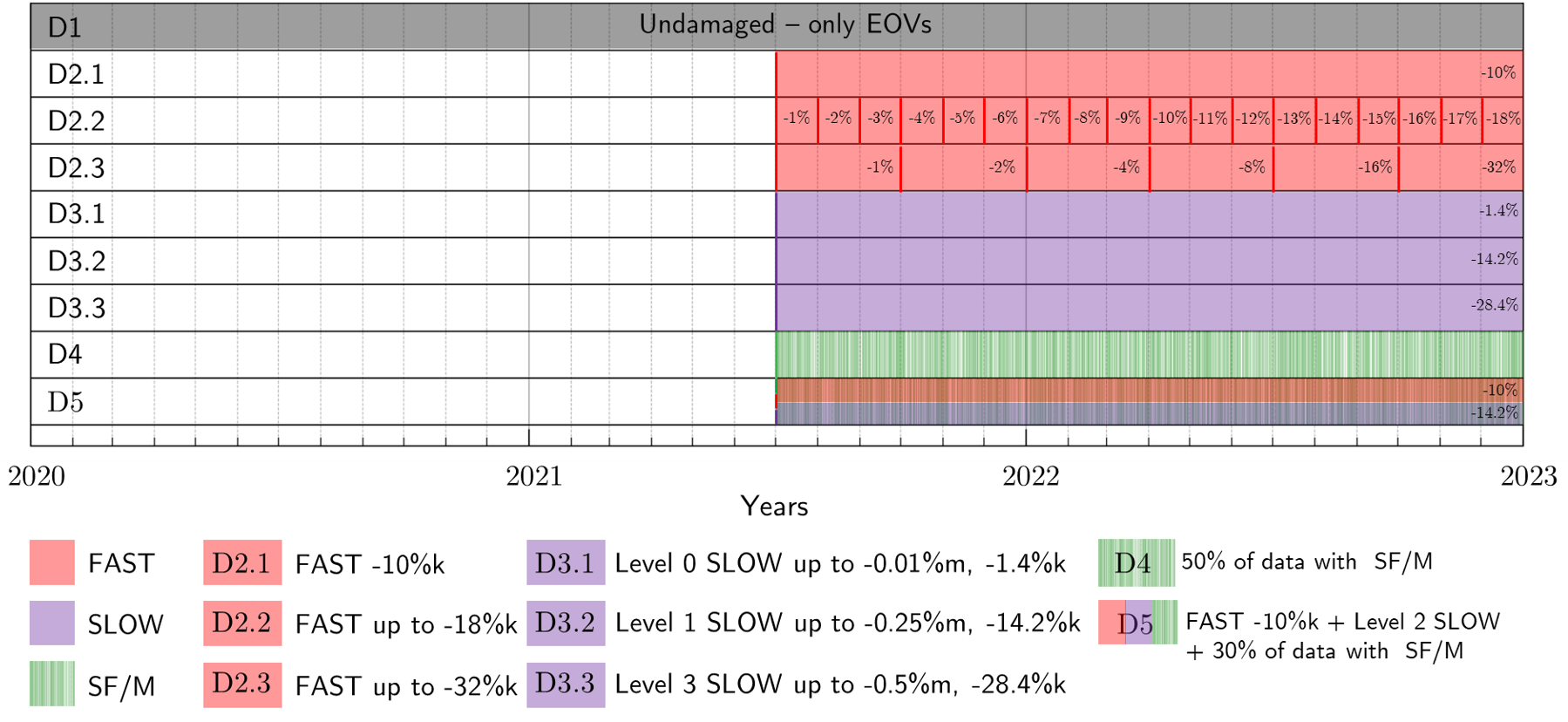}}
    {\caption{Qualitative representation of the sub-datasets content (in red FAST, in purple SLOW, in green SF/M)}\label{fig:Figure12}}
\end{figure}

The full generated dataset consists of 9 sub-datasets of both dynamic and static synthetic acquisitions (see Table~\ref{tab:generated_datasets} and Figure~\ref{fig:Figure12}). All data and scripts are contained in a database at the provided link in the data availability statement. The reference simulation time covers a period of three years (e.g., 2020-2022), for each sampled hour, a simulated acceleration response was generated in the form of three-minute acceleration records with 100 samples per second, together with a single displacement value. The acceleration signals are provided in HDF5 (\texttt{.h5}) format, expressed in $\mathrm{m}/\mathrm{s}^2$. 

Each acceleration file is named using the convention \texttt{accXXXXX-YZ.h5}, where \texttt{XXXXX} is a five-digit, zero-padded numerical code corresponding to the fixed 3-year time series, $Y$ varies from 0 to 5 according to the sub-dataset to which the signal belongs, and $Z$ corresponds to a subset alternative when present. Sub-datasets 2 and 3 both present three alternatives, simulating different levels of fast- and slow-varying damage; each alternative is indicated by adding an increasing number to the dataset index (i.e., \texttt{acc00000-21.h5} belongs to sub-dataset 2, alternative 1). To import each \texttt{.h5} file, the denomination \texttt{acc} must be used.

Deflection acquisitions, expressed in $\mathrm{mm}$, are provided in a single text file (\texttt{deflection\_D1-5.txt}), arranged as a matrix, each column corresponding to sub-dataset $Y$. Each column of the matrix represents a sub-dataset (in this case the data generated for the reference initial period is repeated for each sub-dataset). SF/M contamination does not alter the deflection data in sub-dataset D4.

The first sub-dataset includes the complete time series with only the effect of EOVs on the undamaged structure. Each subsequent sub-dataset is generated as a portion of the initial undamaged dataset to include one or more of the following effects:
\begin{itemize}
\item The effect of fast-varying damage (FAST) acting suddenly on the data (shift), modelled as one or more sudden stiffness's ($k$) decays.
\item The effect of slow-varying damage (SLOW) acting progressively on the data (trend), simulating accelerated corrosion conditions at different rates, decreasing the area of the beam and therefore its stiffness ($k$) and mass ($m$).
\item The effect of sensor faults and/or malfunctions (SF/M) applied \emph{a posteriori} to the simulated acceleration data.
\end{itemize}

\begin{table}[t]
\caption{Generated datasets}
\label{tab:generated_datasets}
\centering
\renewcommand{\arraystretch}{1.3}
\setlength{\tabcolsep}{4pt}

\begin{tabular}{p{0.8cm} p{2.6cm} p{4cm} p{2.8cm} p{1.4cm} p{1cm}}
\toprule
\textbf{Code} & \textbf{Description} &
\textbf{Simulated effects} & \textbf{Duration} &
\textbf{Num. Acq.} & \textbf{Dim. [GB]} \\
\midrule

D1 & Undamaged (UD) &
EOVs &
3 years\par (01/01/2020-\par31/12/2022) &
26280 & 1.90 \\

\midrule

D2.1 &
\multirow{3}{2.6cm}{Fast varying damage (FAST)} &
1 single step -10\% (max 10\%) &
\multirow{3}{2.8cm}{1.5 years from index 13104 (01/07/2021--31/12/2022)} &
13176 & 0.98 \\

D2.2 & &
18 steps of -1\% up to -18\% each month &
 &
13176 & 0.98 \\

D2.3 & &
6 steps of increasing intensity (from 1\% up to 32\%, with geometric progression doubling by 2) &
 &
13176 & 0.98 \\

\midrule

D3.1 &
\multirow{3}{2.6cm}{Slow varying damage (SLOW)} &
Ageing damage (Lev. 0) &
\multirow{3}{2.8cm}{1.5 years from index 13104 (01/07/2021--31/12/2022)} &
13176 & 0.98 \\

D3.2 & &
Accelerated \par corrosion (Lev. 1) &
 &
13176 & 0.98 \\

D3.3 & &
Accelerated \par corrosion (Lev. 3) &
 &
13176 & 0.98 \\

\midrule

D4 &
Sensor Faults  \par and/or  \par Malfunctions (SF/M) &
$\leq 50\%$ of contaminated data equally distributed among 7 classes of sensor faults or malfunctions &
6600 acquisitions randomly sampled in the period from index 13104 to the end &
6600 & 0.49 \\

\midrule

D5 &
Overlap &
FAST of dataset 2.3 and SLOW of dataset 3.3 overlapped with 30\% of the data contaminated with sensor faults or malfunctions &
1.5 years from index 13104 (01/07/2021-\par31/12/2022) &
13176 & 0.98 \\

\bottomrule
\end{tabular}
\end{table}

\begin{figure}[!h]
    \FIG{\includegraphics[width = 0.6\linewidth]{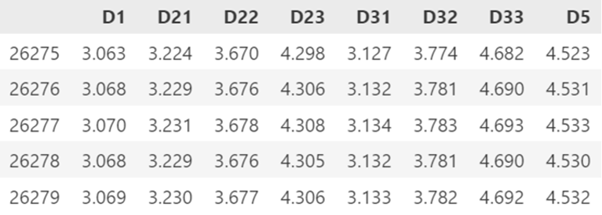}}
    {\caption{Excerpt of the tail of the deflection dataset; sample IDs in the first column, deflection values are in $\mathrm{mm}$}\label{fig:Figure13}}
\end{figure}

The time-varying loading and temperature profiles used are provided, together with the series of $k$-$m$ pairs used to generate each sub-dataset (all in \texttt{.txt} format in the folder \texttt{input}). Finally, 9 \texttt{.txt} files containing labels for all datasets are provided (see Table~\ref{tab:generated_datasets}). An example of an excerpt of the deflection dataset containing all deflection data is shown in Figure~\ref{fig:Figure13}.

The code is provided in the form of four Jupyter Notebooks written for Python v.~3.11.9 containing the following information:
\begin{itemize}
\item Description of the definition of the EOVs applied (\texttt{EOVs.ipynb}).
\item SDOF system modelling (\texttt{SDOF.ipynb}).
\item Data generation script example (\texttt{SHM.ipynb}).
\item Sensor faults/malfunctions formulation and dataset contamination (\texttt{SFM.ipynb}).
\end{itemize}

Each notebook is documented to guide users in modifying simulations to generate new data or adjust assumed parameters. The data is ready for use \emph{as is} or can be tailored using the provided code to explore different simulation scenarios, for example, simulating damage shifts at different indices or extending the monitoring period. The entire database occupies approximately 10\,GB, encompassing all files from all sub-datasets for the three-year monitoring period. The database therefore includes the generated data, all all necessary input data for the generation, and the code required to operate the simulation.

\vspace{-22pt}
\section{Conclusions and future work}
This paper presented the design and development of a controlled, multi-year numerical SHM benchmark combining Environmental and Operational Variabilities (EOVs), fast- and slow-varying structural changes, and Sensor Faults and Malfunctions (SF/M) within the same dataset. The case study consists of a fixed-fixed steel beam represented through Euler-Bernoulli theory for the static response and an equivalent SDOF model for the dynamic response. Three years of hourly static deflection measurements and three-minute ambient-vibration acceleration acquisitions were generated, and the resulting data were organised into nine sub-datasets in which the different simulated effects occur individually or in combination. The main contribution of the benchmark is to provide a controlled and fully labelled SHM dataset in which several sources of variability that commonly coexist in long-term monitoring can be investigated separately and jointly. The numerical formulation provides direct knowledge of the imposed structural, environmental, operational, and measurement-related changes, while the paired static and dynamic responses allow their effects on different types of monitoring data to be studied under the same conditions. This provides a reproducible baseline for the development and comparison of Damage-Sensitive Features (DSFs), EOV compensation and disentanglement methods, anomaly and damage detection approaches, sensor-fault identification, and statistical, Machine Learning (ML), and Deep Learning (DL) methods. The intended applicability and the limitations associated with the adopted modelling assumptions were discussed in Section~2.2.
	
The dataset is already being used as a test-bed for further research on DSFs and damage identification \cite{marafiniLongTermAgeingDamage2025}. Future developments can progressively extend the benchmark in several directions, including:
	
	\begin{itemize}
		\item extending the equivalent SDOF representation to an MDOF structural model and simulating sensor arrays, allowing the use of multiple modes, mode shapes, modal flexibility, spatial information, and damage-localisation approaches;
		\item introducing nonlinear structural-response mechanisms, including contact or slip at connections and supports, crack opening and closing;
		\item refining the representation of environmental variability through spatial and through-section temperature gradients, restrained thermal expansion, coupled thermo-mechanical effects, and additional environmental covariates;
		\item introducing richer operational and ambient-excitation models calibrated against measured wind, micro-tremors, and other sources;
		\item extending the SF/M classification to include further monitoring-system effects, such as changes in sensor mounting conditions, maintenance interventions, or other anomalies dependent on hardware conditions;
		\item comparing selected simulated data with suitable experimental data;
		\item introducing a broader probabilistic treatment of uncertainties in geometry, material properties, boundary conditions, loading, and model parameters.
	\end{itemize}
	
These extensions would increase the physical complexity and range of methods that can be evaluated while retaining the main objective of the benchmark: providing openly accessible, reproducible, and controlled data for the systematic assessment of SHM methodologies.

\begin{Backmatter}
\paragraph{Funding Statement}
The research was conducted during the course of a PhD International program at the University of Florence supported by a scholarship granted by Regione Toscana under the "Giovanisì" project, in cotutelle with the University of Minho and the Institute , with the additional support of the Erasmus + Traineeship Program.
\paragraph{Competing Interests}
The authors declare that they have no known competing financial interests or personal relationships that could have appeared to influence the work reported in this paper.
\paragraph{Data Availability Statement}
The dataset generated and the code used for its definition are publicly available through the Zenodo repository under the name SDOF SHM benchmark. The data can be accessed via the persistent identifier \url{https://doi.org/10.5281/zenodo.17900300}{}. The repository contains the full dataset required to reproduce the generate the data presented in this paper. The simulations were performed on servers at the University of Florence and the University of Minho.
\paragraph{AI Use Statement}
The authors used an AI language model during the study preparation process to assist with text and code editing, proofreading and formatting. AI was not used to generate scientific results, analyses, or interpretations. All content was reviewed and validated by the author.
\paragraph{Ethical Standards}
The research meets all ethical guidelines, including adherence to the legal requirements of the study country.
\paragraph{Authors' Contributions}
Francesca Marafini: Conceptualization, Methodology, Software, Formal analysis, Investigation, Data Curation, Writing Original Draft, Writing Review and Editing, Visualization, Resources; Giacomo Zini: Conceptualization, Validation, Supervision, Writing - Review and Editing, Resources; Alberto Barontini: Conceptualization, Validation, Supervision, Writing - Review and Editing, Resources; Nuno Mendes: Supervision, Writing Review and Editing, Project Administration; Alice Cicirello: Conceptualization, Supervision, Writing Review and Editing; Michele Betti: Conceptualization, Supervision, Project Administration, Funding acquisition - Gianni Bartoli: Conceptualization, Supervision
\bibliography{DiB}
\end{Backmatter}
\end{document}